\documentclass[10pt,twocolumn]{article}
\usepackage{geometry}
\usepackage{titlesec}
\titleformat{\section}{\normalfont\fontsize{10}{12}\bfseries}{\thesection}{1em}{}
\titleformat{\subsection}{\normalfont\fontsize{10}{12}\bfseries}{\thesubsection}{1em}{}
\usepackage[utf8]{inputenc}
\usepackage{graphicx}
\usepackage{dcolumn}
\usepackage{amsmath}
\usepackage{graphicx}
\usepackage[figurename=Fig.,labelfont=bf,labelsep=space,font=small]{caption}
\usepackage[dvipsnames]{xcolor}
\usepackage{hyperref}
\usepackage{booktabs}
\usepackage{authblk}
\usepackage{amsfonts}
\usepackage[font={small},labelfont={bf}]{caption}
\usepackage{color,soul}
\renewcommand{\thesection}{\arabic{section}}

\renewcommand{\thesubsection}{\arabic{section}.\arabic{subsection}}
\renewcommand{\thefigure}{\arabic{figure}}
\renewcommand{\thetable}{\arabic{table}}
\hypersetup{%
  colorlinks,
  plainpages=false,
  breaklinks,
  pdfview=Fit,
  bookmarksopen,
  bookmarksnumbered,
  linkcolor=blue,
  anchorcolor=black,
  citecolor=blue,
  filecolor=black,
  }

\title{\textbf{\textcolor{blue}{The universality of filamentation-caused challenges of ultrafast laser energy deposition in semiconductors}}}

\author[1,*]{Maxime~Chambonneau}
\author[1]{Markus~Blothe}
\author[2]{Vladimir~Yu.~Fedorov}
\author[3]{Isaure~de~Kernier}
\author[4,5]{Stelios~Tzortzakis}
\author[1,6]{Stefan~Nolte}

\affil[1]{Friedrich Schiller University Jena, Institute of Applied Physics, Abbe Center of Photonics, Albert-Einstein-Straße 15, 07745 Jena, Germany}
\affil[2]{Laboratoire Hubert Curien, Université Jean Monnet, Saint-Etienne, France}
\affil[3]{First Light Imaging S.A.S., Europarc Ste Victoire Bât 5, Route de Valbrillant, 13590 Meyreuil, France}
\affil[4]{Institute of Electronic Structure and Laser (IESL), Foundation for Research and Technology—Hellas (FORTH), P.O. Box 1527, GR-71110 Heraklion, Greece}
\affil[5]{Materials Science and Technology Department, University of Crete, 71003 Heraklion, Greece}
\affil[6]{Fraunhofer Institute for Applied Optics and Precision Engineering IOF, Center of Excellence in Photonics, Albert-Einstein-Straße 7, 07745 Jena, Germany}
\affil[*]{\href{mailto:maxime.chambonneau@uni-jena.de}{\textcolor{blue}{maxime.chambonneau@uni-jena.de}}, \href{mailto:maxime.chambonneau@hotmail.fr}{\textcolor{blue}{maxime.chambonneau@hotmail.fr}}}

\date{}

\graphicspath{{Figs/}}

\begin{document}

\twocolumn[\begin{@twocolumnfalse}
\maketitle

\begin{abstract}
Light propagation in semiconductors is the cornerstone of emerging disruptive technologies holding considerable potential to revolutionize telecommunications, sensors, quantum engineering, healthcare, and artificial intelligence. Sky-high optical nonlinearities make these materials ideal platforms for photonic integrated circuits. The fabrication of such complex devices could greatly benefit from in-volume ultrafast laser writing for monolithic and contactless integration. Ironically, as exemplified for Si, nonlinearities act as an efficient immune system self-protecting the material from internal permanent modifications that ultrashort laser pulses could potentially produce. While nonlinear propagation of high-intensity ultrashort laser pulses has been extensively investigated in Si, other semiconductors remain uncharted. In this work, we demonstrate that filamentation universally dictates ultrashort laser pulse propagation in various semiconductors. The effective key nonlinear parameters obtained strongly differ from standard measurements with low-intensity pulses. Furthermore, the temporal scaling laws for these key parameters are extracted. Temporal-spectral shaping is finally proposed to optimize energy deposition inside semiconductors. The whole set of results lays the foundations for future improvements, up to the point where semiconductors can be selectively tailored internally by ultrafast laser writing, thus leading to countless applications for in-chip processing and functionalization, and opening new markets in various sectors including technology, photonics, and semiconductors.\\

\end{abstract}
\end{@twocolumnfalse}]

\begin{figure}[!ht]
\centering
\includegraphics[width=\linewidth]{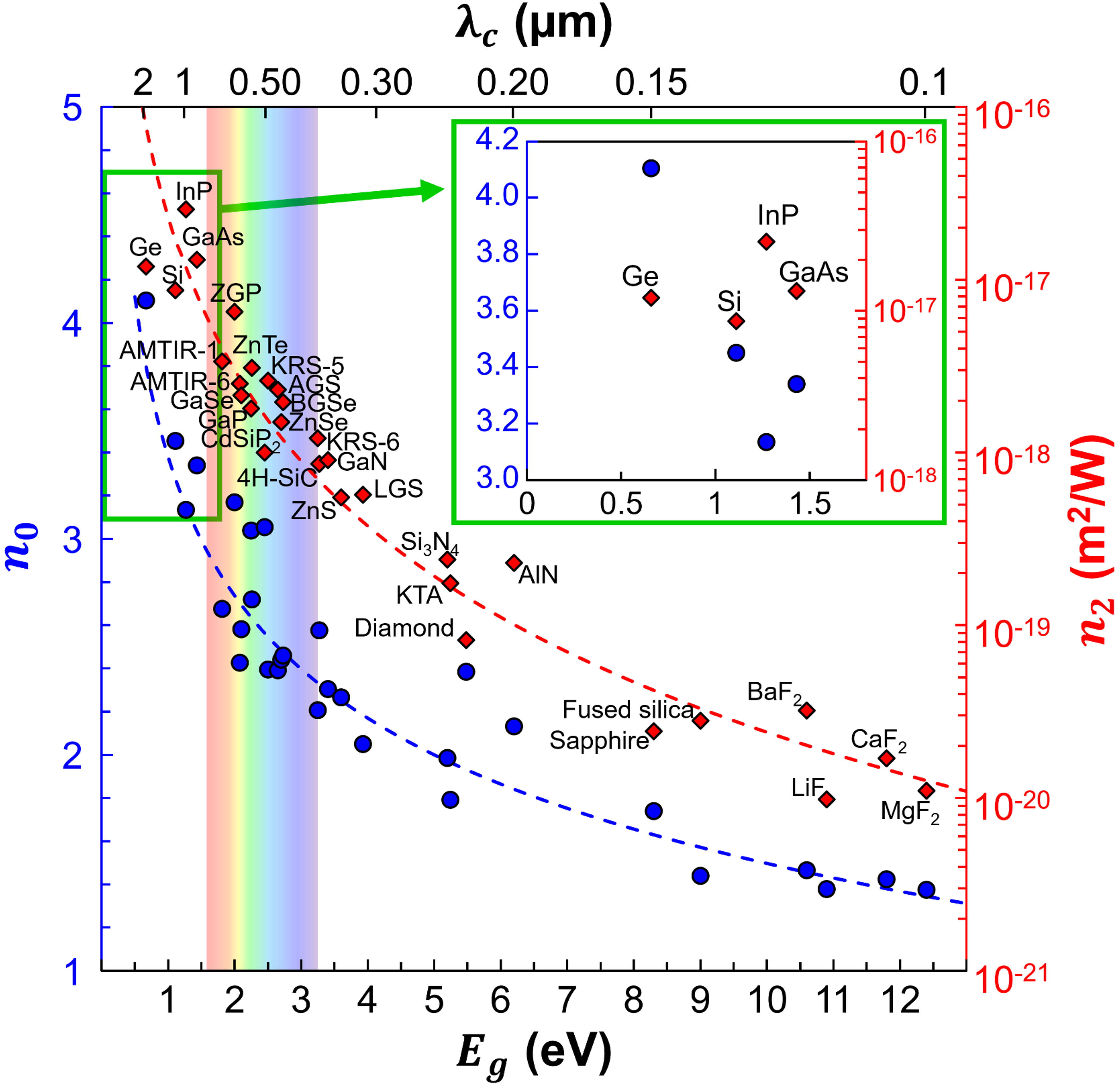}
\caption{\label{fig:Fig1} Dependence of the linear ($n_{0}$, in blue, linear scale) and nonlinear ($n_{2}$, in red, logarithmic scale) refractive index of 30 materials on their band gap ($E_{g}$, corresponding to a cutoff wavelength $\lambda_{c}=hc/E_{g}$, where $h$ and $c$ are the Planck constant and the speed of light in vacuum, respectively). The data given at a wavelength of $\lambda=1960$~nm or close to are extracted from the references indicated in Supplementary Information, Section~\ref{sec:OptParam}. The semiconductors explored in this study are highlighted in the inset.}
\end{figure}

Ultrafast laser filamentation is an extremely nonlinear propagation regime where nonlinear refraction competes with plasma effects \cite{Couairon2007,Berge2007}. In gases, the remarkable properties of filaments have led to a plethora of applications including light detection and ranging (LIDAR) \cite{Kasparian2003}, spectroscopy \cite{Stelmaszczyk2004}, terahertz wave generation \cite{Fedorov2020}, fog clearing \cite{Schimmel2018}, air waveguides \cite{Fu2022,Goffin2023}, and lightning guidance \cite{Houard2023}. In wide-gap solids, the much higher nonlinear refractive index can be advantageously exploited for supercontinuum generation \cite{Alfano1970,Brodeur1999}, laser direct writing of elongated structures \cite{Sudrie2002} and fiber Bragg gratings \cite{Rahnama2020}. However, in narrow-gap materials such as semiconductors, the understanding of ultrafast filamentation is to date limited to Si. In contrast with other media, nonlinear propagation effects in Si are disastrous when aiming for internal structuring, as the energy deposition is delocalized and saturates below the modification threshold due to intensity clamping \cite{Kononenko2012,Kononenko2016,Zavedeev2016,Fedorov2016,Mareev2020,Chambonneau2021a,Grojo2023}. Thanks to characterizations of nonlinear propagation of femtosecond laser pulses in Si, various circumvention techniques have been devised for modifying the bulk of Si, by exploiting surface seeds \cite{Alberucci2020,Chambonneau2023}, hyper numerical aperture \cite{Sreevinas2012,Chanal2017}, longer pulses in the picosecond \cite{Chambonneau2019a,Mareev2020,Das2020,Blothe2024} and nanosecond \cite{Tokel2017,AsgariSabet2024} regime, pulse trains \cite{Wang2020a}, and mid-infrared pulses \cite{Mareev2022}. Nevertheless, nothing indicates that the conclusions on filamentation in Si hold in other semiconductors. From the literature (Fig.~\ref{fig:Fig1}), extreme nonlinear refraction is expected in these narrow-gap media. This suggests that filaments would form at modest laser pulse energies. Moreover, nonlinear absorption is exalted for narrow band gaps \cite{Wherrett1984,VanStryland1985}, which could indicate that, analogously to Si, prefocal absorption hinders localized energy deposition in the focal region.\\

In this article, we demonstrate that ultrafast laser filamentation dictates energy deposition in narrow-gap semiconductors. Two indirect (Si and Ge) and two direct (InP and GaAs) band-gap semiconductors have been selected (see the inset in Fig.~\ref{fig:Fig1}). Because of their narrow band gaps ($E_{g}<1.5$~eV), these media exhibit high linear and nonlinear refraction ($n_{0}>3$, and $n_{2}>10^{-18}$~m$^{2}$/W, respectively). To work in their transparency spectral range, we employ ultrashort laser pulses at a wavelength of $\lambda=1960$~nm (photon energy of 0.63~eV). This corresponds to the 2-photon absorption (2PA) regime for Si and Ge, and the 3-photon absorption (3PA) regime for InP and GaAs. We observe that filamentation governs the interaction in all tested semiconductors. We exploit the reduced energy deposition with ultrashort pulses below the modification threshold to determine the three-dimensional (3D) fluence distribution in these materials, which eventually leads us to define key nonlinear interaction parameters, including the peak fluence $F_{p}$, the effective critical power $P_{\rm{cr}}^{\rm{eff}}$, the effective 2- and 3-photon absorption coefficients $\beta_{2}^{\rm{eff}}$ and $\beta_{3}^{\rm{eff}}$, the fraction of absorbed energy $f_{E}$, and the characteristic absorption length $L_{\rm{abs}}$. Repeating the measurements for pulse durations $\tau=275$~fs~--~25~ps, the temporal scaling laws for these parameters are determined. Ultimately, we propose temporal-spectral shaping approaches to increase energy deposition inside semiconductors.

\section*{A shared filamentation behavior}

\begin{figure}[!ht]
\centering
\includegraphics[width=\linewidth]{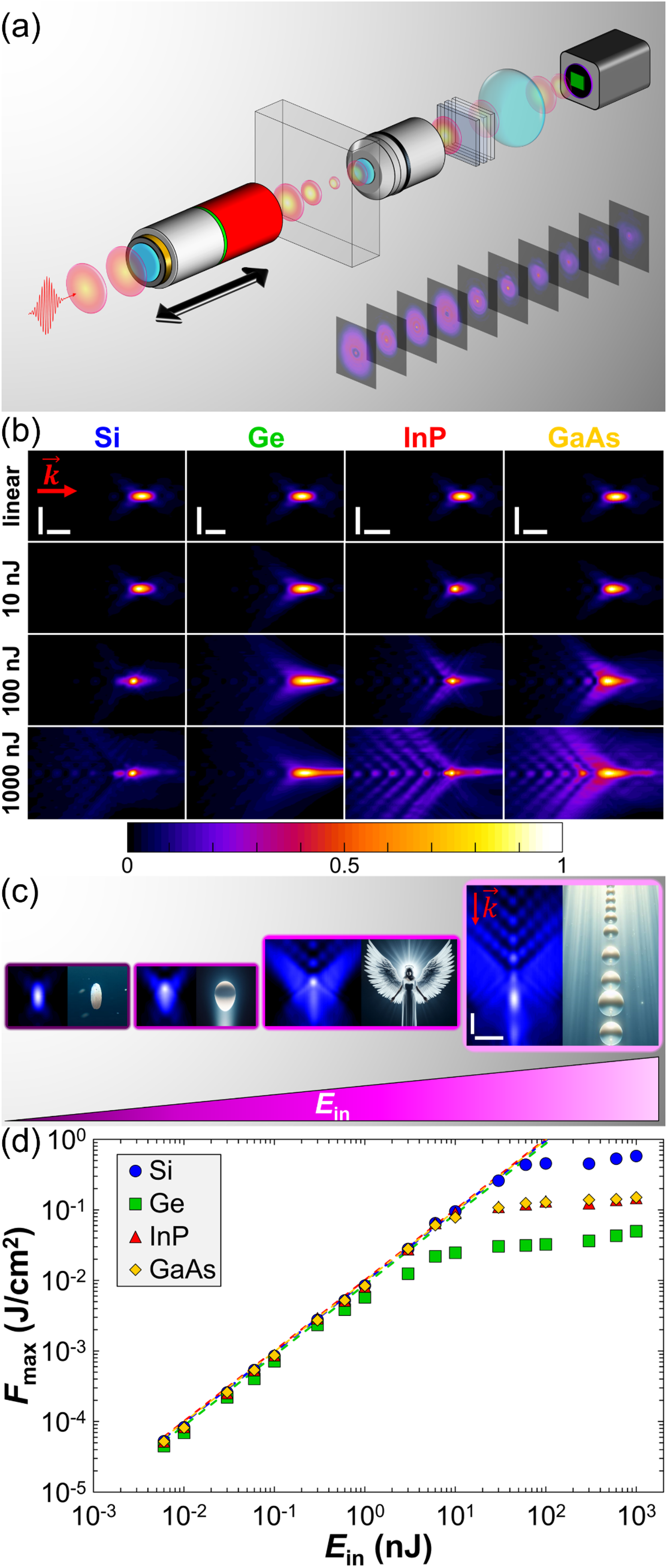}
\caption{\label{fig:Fig2}
(a) Schematic of the nonlinear propagation imaging set-up. (b) Normalized fluence distributions for various $E_{\rm{in}}$ (pulse duration: $\tau=900$~fs). The vector $\vec{k}$ indicates the direction of propagation. (c) Evolutive filament morphology with $E_{\rm{in}}$. The fluence distributions have been recorded in Ge for $\tau=10$~ps. (d) Evolution of the internal maximum fluence $F_{\rm{max}}$ as a function of $E_{\rm{in}}$ for different semiconductors ($\tau=900$~fs). The dashed lines correspond to the linear propagation regime. The radial and on-axis scale bars in (b) and (c) are 10~$\mu$m and 100~$\mu$m, respectively, and apply to all images for the same material.}
\end{figure}

Our approach to characterize filamentation in semiconductors relies on nonlinear propagation imaging. This technique, analogous to tomography as illustrated in Fig.~\ref{fig:Fig2}(a), was initially developed for characterizing light propagation in water \cite{Jarnac2014}, and later applied to dielectrics \cite{Xie2015,Ardaneh2022}. Concerning semiconductors, it was successfully employed to examine nonlinear propagation of light in Si \cite{Fedorov2016,Chanal2017,Mareev2020,Wang2020a,Chambonneau2021,Chambonneau2023}, but it has been so far limited to this material. As shown by the fluence distributions in Fig.~\ref{fig:Fig2}(b), nonlinear propagation imaging is also applicable to other semiconductors. It is all the more favored by the fact that the modification threshold is not crossed (see Supplementary Information, Section~\ref{sec:metrology}). For each material, increasing the input pulse energy $E_{\rm{in}}$ qualitatively results in a spatially extended focal zone with respect to the linear regime. This is ascribable to the competition between Kerr and plasma effects, i.e., the formation of a filament. Quantitatively, the fluence distribution for identical laser conditions strongly depends on the intrinsic nonlinear refraction and absorption properties of the considered medium. Nevertheless, a common feature between all materials is the $E_{\rm{in}}$-dependent evolutive morphology of the fluence distribution, as exemplified in Fig.~\ref{fig:Fig2}(c). For low $E_{\rm{in}}$ values, the propagation is linear and the resulting fluence distribution takes the form of a \textit{grain of rice}. The symmetry of this shape is broken for increased $E_{\rm{in}}$, and distortions toward the prefocal region are observed. This \textit{egg} morphology originates from the Kerr effect, which redistributes the fluence. When $E_{\rm{in}}$ is further increased, prefocal absorption gives rise to the formation of wings, and an \textit{angel} morphology appears. This morphology eventually breaks up into multiple foci for the highest $E_{\rm{in}}$. This \textit{pearl necklace} morphology highlights the complex focusing and defocusing dynamics of filamentation. The evolutive morphology in Fig.~\ref{fig:Fig2}(c) has been observed for all tested semiconductors and pulse durations (see Supplementary Information, Section~\ref{ssec:morpho}).\\

Among all interaction parameters that can be extracted from the 3D fluence distributions (see hereafter), a low-hanging fruit is the maximum fluence $F_{\rm{max}}$. As shown in Fig.~\ref{fig:Fig2}(d) where $\tau=900$~fs, in all semiconductors, $F_{\rm{max}}$ scales linearly with $E_{\rm{in}}$ for sub-nJ values. This is in excellent agreement with theoretical predictions in the linear regime (see Supplementary Information, Section~\ref{sec:linpropregime}). In contrast, when $E_{\rm{in}}$ exceeds a medium-dependent threshold, the experimental data deviate from this regime, and $F_{\rm{max}}$ saturates to a peak value $F_{p}$, which also depends on the material. This behavior is a direct consequence of intensity clamping, which is a typical feature of filamentation \cite{Couairon2007,Berge2007}. This implies that increasing the deposited energy in semiconductors by simply increasing $E_{\rm{in}}$ is a strategy doomed to failure.

\section*{Temporal scaling laws}

\begin{figure*}[!ht]
\centering
\includegraphics[width=\linewidth]{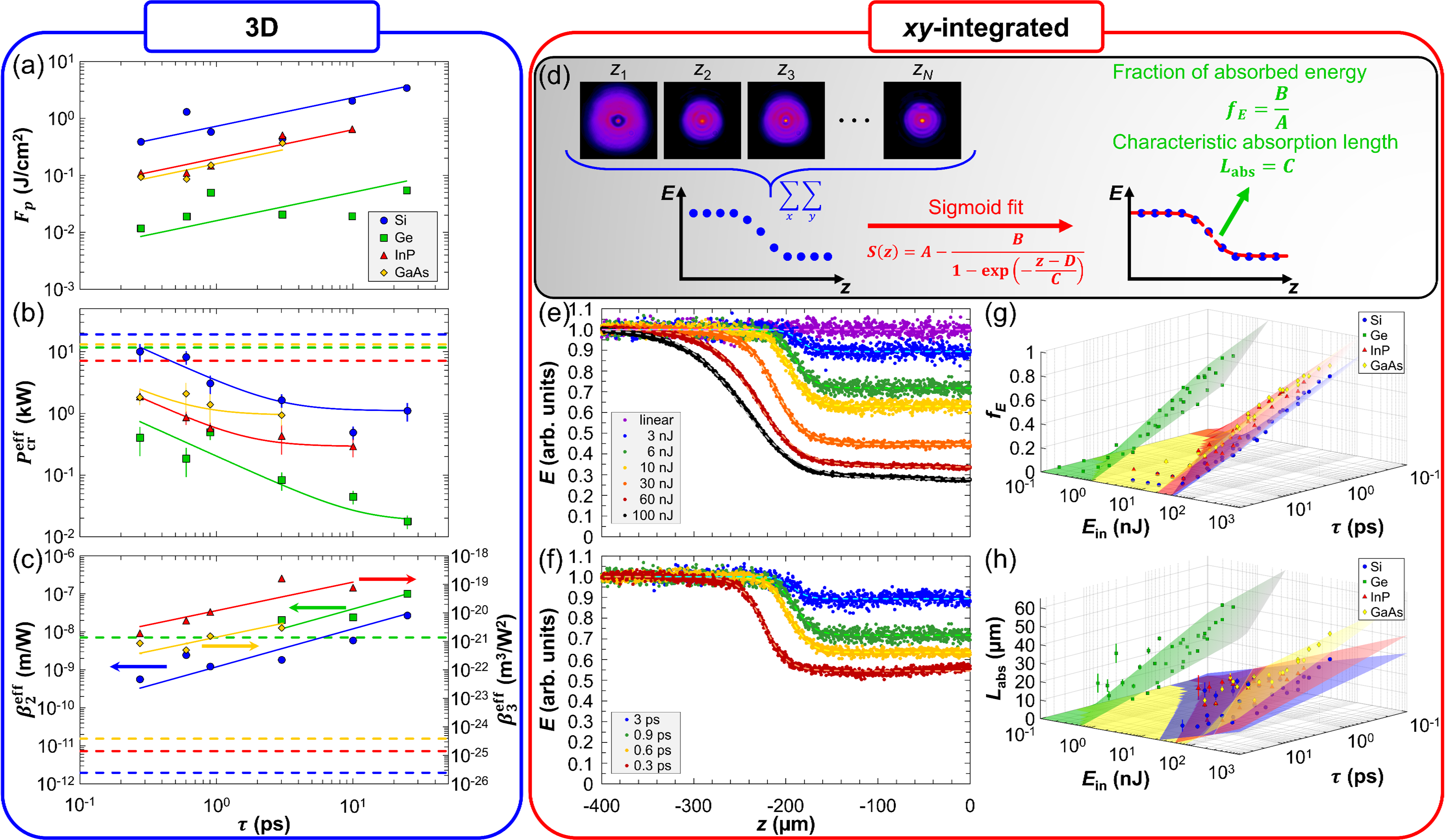}
\caption{\label{fig:Fig3} Temporal scaling laws for key nonlinear interaction parameters in semiconductors. Evolution of (a) the peak fluence $F_{p}$, (b) the effective critical power $P_{\rm{cr}}^{\rm{eff}}$, and (c) the effective 2PA and 3PA coefficients $\beta_{2}^{\rm{eff}}$ (for Si and Ge) and $\beta_{3}^{\rm{eff}}$ (for InP and GaAs) as a function of the pulse duration $\tau$. The theoretical approach for determining $P_{\rm{cr}}^{\rm{eff}}$, $\beta_{2}^{\rm{eff}}$, and $\beta_{3}^{\rm{eff}}$ is given in Supplementary Information, Section~\ref{sec:filamentationregime}. The solid lines in (a) and (c) are $\sqrt{\tau}$ and linear fits, respectively. The solid lines in (b) are calculated from the model described in Supplementary Information, Section~\ref{sec:filamentationregime}, where a Gaussian medium response function is considered. The dashed lines in (b) and (c) correspond to literature values given in Supplementary Information, Table~\ref{tab:materials}. (d) Description of the transversely integrated method to determine and fit the on-axis energy profiles $E(z)$. Mathematical details are given in Supplementary Information, Section~\ref{sec:filamentationregime}. Examples of $E(z)$ profiles in GaAs are shown for various (e) input pulse energies with $\tau=0.9$~ps, and (f) pulse durations with $E_{\rm{in}}=6$~nJ. Pulse duration and input pulse energy dependence of (g) the fraction of absorbed energy $f_{E}$, and (h) the characteristic absorption length $L_{\rm{abs}}$ for all tested semiconductors. The data points are experimental values, and the corresponding surfaces are logarithmic fits, as detailed in Supplementary Information, Section~\ref{sec:filamentationregime}. The experimental error on $f_{E}$ is $<10\%$. Supplementary Videos~1 and 2 offer rotating visualization of the 3D plots in (g) and (h), respectively.}
\end{figure*}

Generally speaking, internal energy deposition strongly depends on the pulse duration, as the laser intensity scales inversely with this parameter. In contrast with wide-gap solids, it was demonstrated that longer pulses lead to higher intensities inside Si \cite{Chambonneau2019a,Mareev2020,Das2020}. This counter-intuitive result originates from decreased propagation nonlinearities when employing longer pulses, resulting in higher peak fluences. As shown in Fig.~\ref{fig:Fig3}(a), this trend is common to all examined semiconductors. Interestingly, in the tested pulse duration range ($\tau=275$~fs~--~25~ps), the peak fluence $F_{p}$ scales as $\sqrt{\tau}$, which implies that increasing $\tau$ by a factor 100 results in an increase of $F_{p}$ by only one order of magnitude. This highlights that, even for the longest pulses employed ($\tau=25$~ps), filamentation still dominates the interaction.\\

Besides the peak fluence $F_{p}$, we also determine key parameters related to nonlinear refraction and absorption. First, the experimental on-axis fluence profiles are compared to linear propagation calculations to extract the effective critical power $P_{\rm{cr}}^{\rm{eff}}$ above which nonlinearities start to alter the propagation (see Supplementary Information, Section~\ref{sec:filamentationregime} for more details on the method). While one could expect $P_{\rm{cr}}^{\rm{eff}}$ to be a constant material property, Fig.~\ref{fig:Fig3}(b) shows that this parameter decreases with the pulse duration for all tested semiconductors. Such a temporal dependence of the effective critical power was observed in air \cite{Ripoche1997,Liu2005} as well as fused silica \cite{Couairon2005}, and explained by the fact that Kerr nonlinearity includes an instantaneous and a delayed medium response---the latter becoming non-negligible for longer pulse durations. Following an analogous approach for semiconductors, we demonstrate that the experimental trends in Fig.~\ref{fig:Fig3}(b) may also originate from delayed medium response (see Supplementary Information, Section~\ref{sec:filamentationregime} for more details).\\

Nonlinear propagation imaging has also been exploited to extract the effective multi-photon absorption coefficient. To do so, the nonlinear focal shift determined experimentally is compared to our recent theoretical approach \cite{Chambonneau2021}, which is based on a modified Marburger formula \cite{Marburger1975}, where power losses are accounted for. The effective critical power values $P_{\rm{cr}}^{\rm{eff}}$ in Fig.~\ref{fig:Fig3}(b) are used as an input parameter for this model. While this approach was initially validated in the case of 2PA in Si, it is also applicable for all semiconductors in different multi-photon absorption regimes (see Supplementary Information, Section~\ref{sec:filamentationregime} for more details). The corresponding effective 2PA and 3PA coefficients $\beta_{2}^{\rm{eff}}$ (for Si and Ge) and $\beta_{3}^{\rm{eff}}$ (for InP and GaAs) both scale linearly with $\tau$ as shown in Fig.~\ref{fig:Fig3}(c).\\

An important result lies in the comparison between $P_{\rm{cr}}^{\rm{eff}}$, $\beta_{2}^{\rm{eff}}$, $\beta_{3}^{\rm{eff}}$, and the corresponding literature values for $P_{\rm{cr}}$, $\beta_{2}$, and $\beta_{3}$ [see dashed lines in Fig.~\ref{fig:Fig3}(b) and (c)]. For both nonlinear refraction and absorption, the parameters deduced from nonlinear propagation imaging are orders of magnitude higher than the literature values. This can be ascribed to the methods traditionally employed for determining these nonlinear coefficients. For instance, in the most standard nonlinear optics technique (z-scan \cite{Sheik-Bahae1990a}), pulses with an energy right above the detection threshold for nonlinearities are loosely focused, which results in a change in transmission of a few percent. Therefore, the measured critical power and multi-photon absorption coefficient are valid for weakly excited media. In contrast, in our nonlinear propagation experiments where materials are strongly ionized, the laser-produced plasma plays a critical role in nonlinear refraction (e.g., scattering, plasma defocusing) and absorption (e.g., free-carrier absorption potentially leading to avalanche ionization).\\

Transverse integration of the fluence allows us to determine how the energy $E$ is absorbed during propagation along the optical axis $z$ [Fig.~\ref{fig:Fig3}(d)]. Examples of normalized $E(z)$ profiles are shown in Fig.~\ref{fig:Fig3}(e) and (f) for different $E_{\rm{in}}$ and $\tau$ in GaAs. For intermediate $E_{\rm{in}}$ values where the filament does not exhibit an \textit{angel} or \textit{pearl necklace} morphology, the experimental data are well-described by a sigmoid function (see Supplementary Information, Section~\ref{sec:filamentationregime} for mathematical details). The sigmoid fits contain two key parameters. The first one is the fraction of absorbed energy $f_{E}$, which is determined by the ratio between the energies before and after the interaction. The second parameter is the characteristic absorption length $L_{\rm{abs}}$, which is inversely proportional to the steepness of the sigmoid. Applying this fitting procedure for all conditions and materials, the dependence of $f_{E}$ and $L_{\rm{abs}}$ on $E_{\rm{in}}$ and $\tau$ [Fig.~\ref{fig:Fig3}(g) and (h), respectively] is obtained. Interestingly, $f_{E}$ and $L_{\rm{abs}}$ both scale logarithmically with $E_{\rm{in}}$ and also with $1/\tau$. These 3D plots show that high input intensity (i.e., high $E_{\rm{in}}$ and short $\tau$) leads to a larger fraction of absorbed energy and an extended absorption zone along $z$. Combined with the saturation of the maximum fluence $F_{\rm{max}}$ shown in Fig.~\ref{fig:Fig2}(d), this result shows that prefocal absorption governs the interaction in all tested semiconductors. Conversely, energy deposition is exalted and more localized for increased pulse durations.

\section*{Temporal-spectral shaping}

Besides this first solution consisting of increasing the pulse duration $\tau$ to improve energy deposition in semiconductors, we have explored two additional methods, both relying on temporal-spectral shaping. The first method that has been examined consists of changing the temporal sequence of spectral components, i.e., the chirp. For the same duration, pulses which are not bandwidth-limited can be up- or down-chirped, i.e., the long wavelengths arrive first or last, respectively. While this parameter plays a major role in high-field physics \cite{Chen2008}, its effect is usually marginal at the moderate intensities typically involved in laser processing. Experimental studies on the surface of semiconductors showed that, in the linear absorption regime, the chirp has a moderate effect \cite{Sun2022} or no effect at all \cite{Louzon2005} on the laser-produced carrier density. Analogous observations were made in the 2- and 5-photon absorption case in Si, where the plasma formation threshold is not significantly affected by the chirp \cite{Mareev2021}.\\

\begin{figure*}[!ht]
\centering
\includegraphics[width=\linewidth]{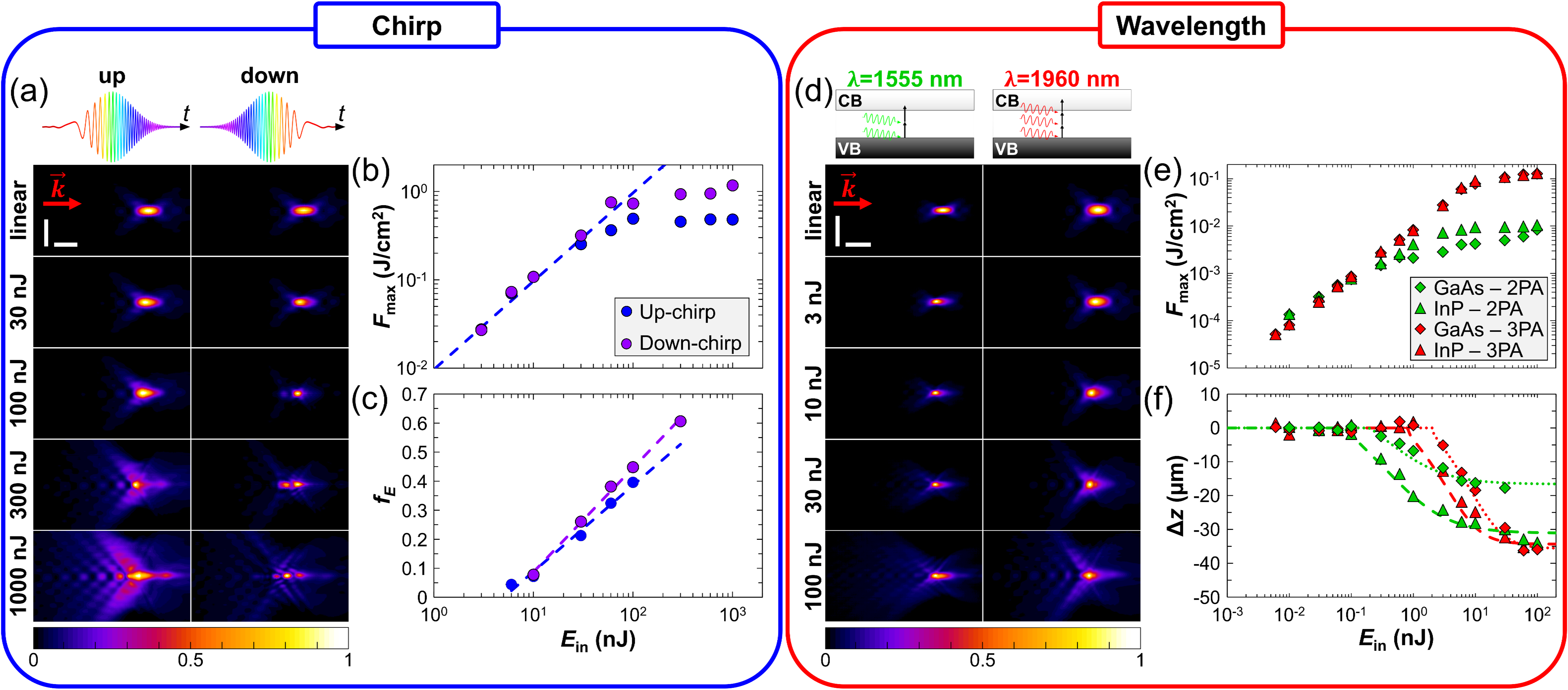}
\caption{\label{fig:Fig4} Temporal-spectral shaping for optimizing energy deposition in semiconductors. The temporal-spectral strategies consist of changing (a)--(c) the chirp, and (d)--(f) the multi-photon absorption order. (a) Normalized fluence distributions in Si for up- and down-chirped 3-ps pulses for various input pulse energies $E_{\rm{in}}$. Evolution of (b) the maximum fluence $F_{\rm{max}}$, and (c) the fraction of absorbed energy $f_{E}$ as a function of $E_{\rm{in}}$ for both chirp configurations. The dashed lines in (b) and (c) correspond to the linear regime, and to logarithmic fits, respectively. (d) Normalized fluence distributions in InP for 1555-nm (2PA) and 1960-nm (3PA) 900-fs pulses for various input pulse energies $E_{\rm{in}}$.  Evolution of (e) the maximum fluence $F_{\rm{max}}$, and (f) the nonlinear focal shift $\Delta z$ as a function of $E_{\rm{in}}$ in InP and GaAs for both wavelengths. The dashed and dotted curves in (f) are calculated with the model described in Supplementary Information, Section~\ref{ssec:MPA}. The vertical and horizontal scale bars are 10~$\mu$m and 100~$\mu$m, respectively, and apply to all images in (a) and (d). The vector $\vec{k}$ indicates the direction of laser propagation.}
\end{figure*}

In most of our experiments, the pulses are up-chirped. In Fig.~\ref{fig:Fig4}(a)--(c), the propagation of 3-ps pulses in Si is displayed for up- and down-chirped pulses. For the same $E_{\rm{in}}$ value, the fluence distributions in Fig.~\ref{fig:Fig4}(a) strongly differ between the two chirps. For down-chirped pulses, energy deposition is much more confined, as confirmed by the $F_{\rm{max}}$ values in Fig.~\ref{fig:Fig4}(b) for $E_{\rm{in}} \ge 60$~nJ. For both chirps, a saturation plateau is reached for $E_{\rm{in}} = 100$~nJ. However, for down-chirped pulses, $F_{p}$ is $2.4 \times$ higher than for up-chirped pulses---thus demonstrating that down-chirped pulses are beneficial for improving internal energy deposition. This is all the more confirmed by the fraction of absorbed energy $f_{E}$ in Fig.~\ref{fig:Fig4}(c), which grows $\approx 20\%$ faster with $E_{\rm{in}}$ for down-chirped pulses in comparison with up-chirped pulses, while the on-axis dimension of the interaction is similar for both chirp configurations [see Fig.~\ref{fig:Fig4}(a)]. Employing the same analysis to determine the effective nonlinear coefficients as in Fig.~\ref{fig:Fig3}, we find the same effective critical power $P_{\rm{cr}}^{\rm{eff}} = 1.6 \pm 0.4$~kW. However, the effective 2PA coefficient differs, with $\beta_{2}=9.2 \times 10^{-10}$~m/W and $1.5 \times 10^{-9}$~m/W for up- and down-chirped pulses, respectively, in agreement with the different nonlinear propagation behaviors in Fig.~\ref{fig:Fig4}(a).\\

The chirp-dependence of filamentation may originate mainly from two physical phenomena. First, the ionization dynamics could differ between up- and down-chirped pulse. Calculations showed that a similar asymmetry exists for fused silica and MgF$_{2}$ \cite{Louzon2005,Sun2022,Duchateau2014}. This is ascribable to the increased efficiency of multi-photon absorption for shorter wavelengths (even for the same multi-photon absorption order \cite{Bristow2007,Lin2007}), while avalanche ionization is more efficient for longer wavelengths. Therefore, in the down-chirped configuration where the ``blue'' spectral components arrive before the ``red'' ones, the produced free-carrier density is higher. The second phenomenon which could play a role is nonlinear dispersion induced by the plasma. Indeed, the interaction results in extended plasma channels along the optical axis, which may show anomalous dispersion, meaning that the ``blue'' can catch up with the ``red'' part of the pulse, in turn resulting in shorter duration---and thus, higher intensity.\\

The second method which has been explored to improve energy deposition inside semiconductors consists of changing the multi-photon absorption order. This can be achieved by selecting different wavelengths corresponding to distinct multi-photon absorption regimes for the same medium. Establishing the optimal wavelength for which in-volume laser--semiconductor interaction is exalted is a problem with no intuitive solutions. Increasing the multi-photon absorption order would theoretically decrease energy deposition at the focus, but also limit prefocal absorption. Conversely, reducing the multi-photon absorption order would have opposite effects. This problem was studied theoretically in Si, where it was predicted that the increased free-carrier absorption efficiency leads to higher energy deposition at longer wavelengths \cite{Zavedeev2016}.\\

To experimentally evaluate the impact of multi-photon absorption order on filamentation, nonlinear propagation imaging has been performed in InP and GaAs for $\tau=900$~fs at a wavelength of $\lambda=1555$~nm (2PA), and then compared to the ones at $\lambda=1960$~nm (3PA) in in Fig.~\ref{fig:Fig4}(d)--(f). As exemplified for InP in Fig.~\ref{fig:Fig4}(d), the wavelength impacts on the fluence distribution, both in the linear and the nonlinear regime. For a given wavelength, an analogous evolution of $F_{\rm{max}}$ for both materials is shown in Fig.~\ref{fig:Fig4}(e) as a function of $E_{\rm{in}}$. This suggests that filamentation governs the interaction both in the 2PA and the 3PA regime. However, an important result is that the peak fluence $F_{p}$ is one order of magnitude higher in the 3PA regime compared to the 2PA regime (maximum ratio of 13 and 21 for InP and GaAs, respectively), suggesting that higher multi-photon absorption orders are beneficial to exalt the interaction.\\

Applying the same approach as for Fig.~\ref{fig:Fig3}(b) and (c), the nonlinear refraction and absorption parameters $P_{\rm{cr}}^{\rm{eff}}$ and $\beta_{2}^{\rm{eff}}$ for $\lambda=1555$~nm are determined. As shown in Fig.~\ref{fig:Fig4}(f), our theoretical approach successfully reproduces the experimental evolution of the nonlinear focal shift $\Delta z$ as a function of $E_{\rm{in}}$ for InP and GaAs at both wavelengths. For both materials, $P_{\rm{cr}}^{\rm{eff}}$ is one order of magnitude lower in the 2PA regime ($57.1 \pm 14.2$~W for InP, $138.1 \pm 69.0$~W for GaAs) compared to the 3PA regime ($573.7 \pm 143.4$~W for InP, and $1387 \pm 693$~W for GaAs). This is ascribable to the normal dispersion of the nonlinear index as well as the $\lambda^{2}$ dependence of the critical power. The data points where $P > P_{\rm{cr}}^{\rm{eff}}$ in Fig.~\ref{fig:Fig4}(f) give access to $\beta_{2}^{\rm{eff}} = 7.8 \times 10^{-8}$~m/W for InP, and $\beta_{2}^{\rm{eff}} = 7.5 \times 10^{-8}$~m/W for GaAs. Similar to the measurement of $\beta_{3}^{\rm{eff}}$ at $\lambda=1960$~nm in Fig.~\ref{fig:Fig3}(c), these $\beta_{2}^{\rm{eff}}$ values are orders of magnitude higher than the $\beta_{2}$ literature values \cite{Vignaud2004,Hurlbut2007,Gonzalez2009,Krishnamurthy2011,Oishi2018}, again attributable to differences in experimental determination method.\\

As a final remark, the three strategies to exalt energy deposition in semiconductors which consist of increasing $\tau$ [Fig.~\ref{fig:Fig3}(a)], employing down-chirped pulses [Fig.~\ref{fig:Fig4}(b)], and selecting $\lambda$ in higher multi-photon absorption order regimes [Fig.~\ref{fig:Fig4}(e)] are not mutually exclusive. To illustrate this aspect, besides the narrow-gap semiconductors investigated in this study, it is interesting to note that the filamentation-caused limitations for energy deposition also persist in wider band-gap semiconductors such as polycrystalline ZnSe ($E_{g}\approx2.7$~eV). Indeed, the threshold for internal modifications in this material could not be crossed with bandwidth-limited pulses with $\tau=115$~fs at $\lambda=800$~nm (2PA regime) \cite{Okhrimchuk2009}. In contrast, by simultaneously using longer pulse durations ($\tau=0.5$~--~2~ps), down-chirped pulses, and $\lambda=1047$~nm (3PA regime), in-volume laser writing was possible \cite{Macdonald2010}. Apart from semiconductors, insulators exhibit even wider band gaps and lower nonlinear refractive index $n_{2}$ (Fig.~\ref{fig:Fig1}). Consequently, the filamentation-caused limitations are drastically reduced in these materials, and internal modifications can be produced in the 2PA regime as exemplified for sapphire \cite{Grivas2008} and fused silica \cite{Saliminia2012}.

\section*{Conclusion}

To summarize, nonlinear propagation imaging has been utilized to demonstrate that filamentation governs ultrafast laser--matter interaction in semiconductors. The 3D fluence distributions obtained exhibit an evolutive morphology, as well as a saturation of the maximum fluence for increased input pulse energy. While, in principle, the impact of propagation nonlinearities would reduce with increased pulse duration, the established temporal scaling laws for effective key nonlinear refraction and absorption parameters show that such a strategy is less beneficial than expected. Nevertheless, for all tested semiconductors, the peak fluence increases with the pulse duration. Other identified strategies for increasing the peak fluence in semiconductors consist of temporal-spectral shaping, which can take the form of down-chirped as well as long wavelength pulses. The whole set of results allows us to envision future optimizations to improve the degree of control of energy deposition in the bulk of semiconductors, with various potential applications including in-chip ultrafast laser functionalization.

\section*{Materials $\&$ methods}

\noindent
\textbf{Samples.} The Si, Ge, InP and GaAs samples used in this study are $500 \pm 25$~$\mu$m thick, $<$100$>$-oriented, undoped, and two-side polished.\\

\noindent
\textbf{Ultrafast laser irradiation.} A Tm-doped fiber laser source (Active Fiber Systems, see details in Ref.~\cite{Baudisch2018}) emitting at a center wavelength of 1960~nm and a repetition rate of 50~kHz has been used. The spectral bandwidth is 29~nm [full width at half maximum (FWHM)]. The pulse duration is measured with autocorrelation and adjustable from 275~fs to 25~ps (FWHM assuming a sech$^{2}$ shape) by detuning the position of gratings in the optical compressor. In all experiments except the ones in Fig.~\ref{fig:Fig4}(a)--(c), the pulses are up-chirped, i.e., long wavelengths arrive before short wavelengths. The input pulse energy is controlled optically by means of a half-wave plate and a linear polarizer, and the beam size before focusing is adjusted with a Galilean telescope. The beam is focused by means of an objective lens of numerical aperture $\rm{NA}=0.40$ (Mitutoyo, M Plan Apo NIR $20 \times$) mounted on a precision linear translation stage (Physik Instrumente, M-404.1DG) allowing its displacement along the optical axis with minimum incremental motion of 100~nm, and a resolution $<12$~nm. For experiments at a wavelength of $\lambda=1555$~nm, an Er-doped fiber laser source (Raydiance Inc., Smart Light 50) delivering non-chirped 900-fs pulses at a repetition rate of 1~kHz was employed. In all experiments the polarization is linear.\\

\noindent
\textbf{Nonlinear propagation imaging.} To record 3D fluence distributions inside semiconductors, nonlinear propagation imaging has been used. This technique is based on an infrared microscope directed opposite to the incoming laser. This microscope is composed of an objective lens of numerical aperture $\rm{NA}=0.85$ (Olympus, LCPLN100XIR), a tube lens (Thorlabs, TTL200-S8), and an extended InGaAs camera (First Light Imaging, C-RED 2 ER 2.2$\mu$m) \cite{DeKernier2022}. This infrared camera is composed of $640 \times 512$ pixels with 15~$\mu$m pixel pitch, and responds linearly in the 1380--2150~nm spectral range. For improved imaging performance, this camera has been operated with a frame rate of 600~Hz, an exposure time of 1.66~ms, and a sensor temperature of $-55^{\circ}$C. For experiments at a wavelength of $\lambda=1555$~nm, a standard InGaAs camera (Xenics, Bobcat 320) composed of $320 \times 256$ pixels, with 20~$\mu$m pixel pitch is used. To optimize the dynamic range of the cameras for all input pulse energies $E_{\rm{in}}$ ranging from 1~pJ to 1~$\mu$J, calibrated neutral density filters have been inserted between the objective lens and the tube lens. The procedure can be divided into three steps. First, the focal plane of the imaging objective is positioned at the exit surface of the sample under white light illumination. Second, at low input pulse energy (typically, a few pJ), the geometrical focus is positioned at the exit surface of the sample, so that it is imaged on the camera. Third, the focusing objective is moved along the optical axis with steps of 100~nm around the geometrical focus position. A typical recording is composed of 2,000 images, with a total stage movement of 200~$\mu$m around the geometrical focus position. The recorded $z$ positions are then multiplied by the linear refractive index of the tested medium to get the actual positions in the material.\\

\noindent
\textbf{Propagation calculations.} To ensure that the 3D fluence distributions obtained for the lowest input pulse energies $E_{\rm{in}}$ correspond to the linear propagation regime, propagation calculations detailed in Supplementary Information, Section~\ref{sec:linpropregime} have been performed with our vectorial model \textit{InFocus} \cite{Li2021,Li2021a}. In this regime, the maximum fluence reads $F_{\rm{max}}=2TE_{\rm{in}}/(\pi w_{0}^{2})$, where $T$ is the medium-dependent Fresnel transmission coefficient at the air--medium interface, and $w_{0}$ is the beam radius at $1/e^{2}$. The corresponding calculations serve as a benchmark for the evaluation of key interaction parameters such as the effective critical power $P_\text{cr}^\text{eff}$ (see Supplementary Information, Section~\ref{ssec:Pcr}), which is in turn implemented in our model based on a modified Marburger formula to determine the effective 2- and 3-photon absorption coefficients $\beta_{2}^{\rm{eff}}$ and $\beta_{3}^{\rm{eff}}$ (see Supplementary Information, Section~\ref{ssec:MPA}).

\section*{Funding}

The German Federal Ministry of Education and Research [Bundesministerium für Bildung und Forschung (BMBF)] through the RUBIN-UKPi\~no project (Grants no.~03RU2U033H and 03RU2U032F).\\

The German Research Foundation [Deutsche Forschungsgemeinschaft (DFG)], through the Silabus (Grant no.~530105422), and the Inseption (Grant no.~545531713) projects.

\section*{Conflict of interest}

The authors declare no conflict of interest.

\section*{Author contributions}

M.C. and M.B. performed the experiments. M.B. automatized the measurements. M.C. processed, analyzed and interpreted the data. V.Y.F., S.T. and M.C. developed the model and performed the numerical simulations. S.N. and M.C. led the project. All authors discussed the results and contributed to the manuscript prepared by M.C.

\bibliographystyle{unsrt}
{\footnotesize
\bibliography{refs}

\begin{thebibliography}{100}

\bibitem{Couairon2007}
A~Couairon and A~Mysyrowicz.
\newblock {Femtosecond filamentation in transparent media}.
\newblock {\em Physics Reports}, 441(2-4):47--189, mar 2007.

\bibitem{Berge2007}
L.~Berg{\'{e}}, S.~Skupin, R.~Nuter, J.~Kasparian, and J-P Wolf.
\newblock {Ultrashort filaments of light in weakly ionized, optically transparent media}.
\newblock {\em Reports on Progress in Physics}, 70(10):1633--1713, oct 2007.

\bibitem{Kasparian2003}
J.~Kasparian, M.~Rodriguez, G.~M{\'{e}}jean, J.~Yu, E.~Salmon, H.~Wille, R.~Bourayou, S.~Frey, Y.-B. Andr{\'{e}}, A.~Mysyrowicz, R.~Sauerbrey, J.-P. Wolf, and W{\"{o}ste, L.}
\newblock {White-Light Filaments for Atmospheric Analysis}.
\newblock {\em Science}, 301(5629):61--64, jul 2003.

\bibitem{Stelmaszczyk2004}
Kamil Stelmaszczyk, Philipp Rohwetter, Guillaume M{\'{e}}jean, Jin Yu, Estelle Salmon, J{\'{e}}r{\^{o}}me Kasparian, Roland Ackermann, Jean-Pierre Wolf, and Ludger W{\"{o}}ste.
\newblock {Long-distance remote laser-induced breakdown spectroscopy using filamentation in air}.
\newblock {\em Applied Physics Letters}, 85(18):3977--3979, 2004.

\bibitem{Fedorov2020}
Vladimir~Yu Fedorov and Stelios Tzortzakis.
\newblock {Powerful terahertz waves from long-wavelength infrared laser filaments}.
\newblock {\em Light: Science $\&$ Applications}, 9(1):186, nov 2020.

\bibitem{Schimmel2018}
Guillaume Schimmel, Thomas Produit, Denis Mongin, J{\'{e}}r{\^{o}}me Kasparian, and Jean-Pierre Wolf.
\newblock {Free space laser telecommunication through fog}.
\newblock {\em Optica}, 5(10):1338, oct 2018.

\bibitem{Fu2022}
Silin Fu, Benoit Mahieu, Andr{\'{e}} Mysyrowicz, and Aurelien Houard.
\newblock {Femtosecond filamentation of optical vortices for the generation of optical air waveguides}.
\newblock {\em Optics Letters}, 47(19):5228, oct 2022.

\bibitem{Goffin2023}
A~Goffin, A~Tartaro, and H~M Milchberg.
\newblock {Quasi-steady-state air waveguide}.
\newblock {\em Optica}, 10(4):505, apr 2023.

\bibitem{Houard2023}
Aur{\'{e}}lien Houard, Pierre Walch, Thomas Produit, Victor Moreno, Benoit Mahieu, Antonio Sunjerga, Clemens Herkommer, Amirhossein Mostajabi, Ugo Andral, Yves-Bernard Andr{\'{e}}, Magali Lozano, Laurent Bizet, Malte~C. Schroeder, Guillaume Schimmel, Michel Moret, Mark Stanley, W.~A. Rison, Oliver Maurice, Bruno Esmiller, Knut Michel, Walter Haas, Thomas Metzger, Marcos Rubinstein, Farhad Rachidi, Vernon Cooray, Andr{\'{e}} Mysyrowicz, J{\'{e}}r{\^{o}}me Kasparian, and Jean-Pierre Wolf.
\newblock {Laser-guided lightning}.
\newblock {\em Nature Photonics}, 17(3):231--235, 2023.

\bibitem{Alfano1970}
R.~R. Alfano and S.~L. Shapiro.
\newblock {Observation of Self-Phase Modulation and Small-Scale Filaments in Crystals and Glasses}.
\newblock {\em Physical Review Letters}, 24(11):592--594, mar 1970.

\bibitem{Brodeur1999}
A.~Brodeur and S.~L. Chin.
\newblock {Ultrafast white-light continuum generation and self-focusing in transparent condensed media}.
\newblock {\em Journal of the Optical Society of America B}, 16(4):637, apr 1999.

\bibitem{Sudrie2002}
L.~Sudrie, A.~Couairon, M.~Franco, B.~Lamouroux, B.~Prade, S.~Tzortzakis, and A.~Mysyrowicz.
\newblock {Femtosecond Laser-Induced Damage and Filamentary Propagation in Fused Silica}.
\newblock {\em Physical Review Letters}, 89(18):186601, oct 2002.

\bibitem{Rahnama2020}
Abdullah Rahnama, Keivan {Mahmoud Aghdami}, Young~Hwan Kim, and Peter~R. Herman.
\newblock {Ultracompact Lens‐Less “Spectrometer in Fiber” Based on Chirped Filament‐Array Gratings}.
\newblock {\em Advanced Photonics Research}, 1(2):2000026, dec 2020.

\bibitem{Kononenko2012}
Vitali~V. Kononenko, Vitali~V. Konov, and Evgeny~M. Dianov.
\newblock {Delocalization of femtosecond radiation in silicon}.
\newblock {\em Optics Letters}, 37(16):3369, aug 2012.

\bibitem{Kononenko2016}
V~V Kononenko, E~V Zavedeev, and V~M Gololobov.
\newblock {The effect of light-induced plasma on propagation of intense fs laser radiation in c-Si}.
\newblock {\em Applied Physics A}, 122(4):293, apr 2016.

\bibitem{Zavedeev2016}
E~V Zavedeev, V~V Kononenko, and V~I Konov.
\newblock {Delocalization of femtosecond laser radiation in crystalline Si in the mid-IR range}.
\newblock {\em Laser Physics}, 26(1):016101, jan 2016.

\bibitem{Fedorov2016}
V.~Yu. Fedorov, M.~Chanal, D.~Grojo, and S.~Tzortzakis.
\newblock {Accessing Extreme Spatiotemporal Localization of High-Power Laser Radiation through Transformation Optics and Scalar Wave Equations}.
\newblock {\em Physical Review Letters}, 117(4):043902, jul 2016.

\bibitem{Mareev2020}
E~I Mareev, K~V Lvov, B~V Rumiantsev, E~A Migal, I~D Novikov, S~Yu Stremoukhov, and F~V Potemkin.
\newblock {Effect of pulse duration on the energy delivery under nonlinear propagation of tightly focused Cr:forsterite laser radiation in bulk silicon}.
\newblock {\em Laser Physics Letters}, 17(1):015402, jan 2020.

\bibitem{Chambonneau2021a}
Maxime Chambonneau, David Grojo, Onur Tokel, Fatih~{\"{O}}mer Ilday, Stelios Tzortzakis, and Stefan Nolte.
\newblock {In‐Volume Laser Direct Writing of Silicon—Challenges and Opportunities}.
\newblock {\em Laser $\&$ Photonics Reviews}, 15(11):2100140, nov 2021.

\bibitem{Grojo2023}
David Grojo, Maxime Chambonneau, Shuting Lei, Alexandros Mouskeftaras, Olivier Ut{\'{e}}za, and Andong Wang.
\newblock {Internal Structuring of Semiconductors with Ultrafast Lasers: Opening a Route to Three-Dimensional Silicon Photonics}.
\newblock In {\em Ultrafast Laser Nanostructuring}, pages 979--1018. Springer, 2023.

\bibitem{Alberucci2020}
Alessandro Alberucci, Namig Alasgarzade, Maxime Chambonneau, Markus Blothe, Helena K{\"{a}}mmer, Gabor Matth{\"{a}}us, Chandroth~P. Jisha, and Stefan Nolte.
\newblock {In-Depth Optical Characterization of Femtosecond-Written Waveguides in Silicon}.
\newblock {\em Physical Review Applied}, 14(2):024078, aug 2020.

\bibitem{Chambonneau2023}
Maxime Chambonneau, Qingfeng Li, Markus Blothe, Stree~Vithya Arumugam, and Stefan Nolte.
\newblock {Ultrafast Laser Welding of Silicon}.
\newblock {\em Advanced Photonics Research}, 4(5):2200300, may 2023.

\bibitem{Sreevinas2012}
V.~V. {Parsi Sreenivas}, M.~B{\"{u}}lters, and R.~B. Bergmann.
\newblock {Microsized subsurface modification of mono-crystalline silicon via non-linear absorption}.
\newblock {\em Journal of the European Optical Society-Rapid Publications}, 7(6):12035, sep 2012.

\bibitem{Chanal2017}
Margaux Chanal, Vladimir~Yu. Fedorov, Maxime Chambonneau, Rapha{\"{e}}l Clady, Stelios Tzortzakis, and David Grojo.
\newblock {Crossing the threshold of ultrafast laser writing in bulk silicon}.
\newblock {\em Nature Communications}, 8(1):773, oct 2017.

\bibitem{Chambonneau2019a}
M.~Chambonneau, L.~Lavoute, D.~Gaponov, V.Y. Fedorov, A.~Hideur, S.~F{\'{e}}vrier, S.~Tzortzakis, O.~Ut{\'{e}}za, and D.~Grojo.
\newblock {Competing Nonlinear Delocalization of Light for Laser Inscription Inside Silicon with a 2-$\mu$m Picosecond Laser}.
\newblock {\em Physical Review Applied}, 12(2):024009, aug 2019.

\bibitem{Das2020}
Amlan Das, Andong Wang, Olivier Uteza, and David Grojo.
\newblock {Pulse-duration dependence of laser-induced modifications inside silicon}.
\newblock {\em Optics Express}, 28(18):26623, aug 2020.

\bibitem{Blothe2024}
Markus Blothe, Alessandro Alberucci, Namig Alasgarzade, Maxime Chambonneau, and Stefan Nolte.
\newblock {Transverse Inscription of Silicon Waveguides by Picosecond Laser Pulses}.
\newblock {\em Laser and Photonics Reviews}, 2400535:1--12, 2024.

\bibitem{Tokel2017}
Onur Tokel, Ahmet Turnalı, Ghaith Makey, Parviz Elahi, Tahir {\c{C}}olakoğlu, Emre Erge{\c{c}}en, {\"{O}}zg{\"{u}}n Yavuz, Ren{\'{e}} H{\"{u}}bner, Mona {Zolfaghari Borra}, Ihor Pavlov, Alpan Bek, Raşit Turan, Denizhan~Koray Kesim, Serhat Tozburun, Serim Ilday, and F.~{\"{O}}mer Ilday.
\newblock {In-chip microstructures and photonic devices fabricated by nonlinear laser lithography deep inside silicon}.
\newblock {\em Nature Photonics}, 11(10):639--645, oct 2017.

\bibitem{AsgariSabet2024}
Rana {Asgari Sabet}, Aqiq Ishraq, Alperen Saltik, Mehmet B{\"{u}}t{\"{u}}n, and Onur Tokel.
\newblock {Laser nanofabrication inside silicon with spatial beam modulation and anisotropic seeding}.
\newblock {\em Nature Communications}, 15(1):5786, jul 2024.

\bibitem{Wang2020a}
Andong Wang, Amlan Das, and David Grojo.
\newblock {Ultrafast Laser Writing Deep inside Silicon with THz-Repetition-Rate Trains of Pulses}.
\newblock {\em Research}, 2020:1--11, jan 2020.

\bibitem{Mareev2022}
Evgenii Mareev, Andrey Pushkin, Ekaterina Migal, Kirill Lvov, Sergey Stremoukhov, and Fedor Potemkin.
\newblock {Single-shot femtosecond bulk micromachining of silicon with mid-IR tightly focused beams}.
\newblock {\em Scientific Reports}, 12(1):7517, may 2022.

\bibitem{Wherrett1984}
B.~S. Wherrett.
\newblock {Scaling rules for multiphoton interband absorption in semiconductors}.
\newblock {\em Journal of the Optical Society of America B}, 1(1):67, mar 1984.

\bibitem{VanStryland1985}
Eric~W. {Van Stryland}, M~A Woodall, H~Vanherzeele, and M~J Soileau.
\newblock {Energy band-gap dependence of two-photon absorption}.
\newblock {\em Optics Letters}, 10(10):490, oct 1985.

\bibitem{Jarnac2014}
Am{\'{e}}lie Jarnac, Gintaras Tamosauskas, Donatas Majus, Aur{\'{e}}lien Houard, Andr{\'{e}} Mysyrowicz, Arnaud Couairon, and Audrius Dubietis.
\newblock {Whole life cycle of femtosecond ultraviolet filaments in water}.
\newblock {\em Physical Review A}, 89(3):033809, mar 2014.

\bibitem{Xie2015}
Chen Xie, Vytautas Jukna, Carles Mili{\'{a}}n, Remo Giust, Ismail Ouadghiri-Idrissi, Tatiana Itina, John~M. Dudley, Arnaud Couairon, and Francois Courvoisier.
\newblock {Tubular filamentation for laser material processing}.
\newblock {\em Scientific Reports}, 5(1):8914, mar 2015.

\bibitem{Ardaneh2022}
Kazem Ardaneh, Remi Meyer, Mostafa Hassan, Remo Giust, Benoit Morel, Arnaud Couairon, Guy Bonnaud, and Francois Courvoisier.
\newblock {Femtosecond laser-induced sub-wavelength plasma inside dielectrics: I. Field enhancement}.
\newblock {\em Physics of Plasmas}, 29(7), jul 2022.

\bibitem{Chambonneau2021}
Maxime Chambonneau, Qingfeng Li, Vladimir~Yu Fedorov, Markus Blothe, Kay Schaarschmidt, Martin Lorenz, Stelios Tzortzakis, and Stefan Nolte.
\newblock {Taming Ultrafast Laser Filaments for Optimized Semiconductor–Metal Welding}.
\newblock {\em Laser $\&$ Photonics Reviews}, 15(2):2000433, feb 2021.

\bibitem{Ripoche1997}
Jean-Francois Ripoche, Georges Grillon, Bernard Prade, Michel Franco, Erik Nibbering, R{\"{u}}diger Lange, and Andr{\'{e}} Mysyrowicz.
\newblock {Determination of the time dependence of n2 in air}.
\newblock {\em Optics Communications}, 135(4-6):310--314, feb 1997.

\bibitem{Liu2005}
W.~Liu and S.~L. Chin.
\newblock {Direct measurement of the critical power of femtosecond Ti:sapphire laser pulse in air}.
\newblock {\em Optics Express}, 13(15):5750, 2005.

\bibitem{Couairon2005}
A.~Couairon, L.~Sudrie, M.~Franco, B.~Prade, and A.~Mysyrowicz.
\newblock {Filamentation and damage in fused silica induced by tightly focused femtosecond laser pulses}.
\newblock {\em Physical Review B}, 71(12):125435, mar 2005.

\bibitem{Marburger1975}
J.H. Marburger.
\newblock {Self-focusing: theory}.
\newblock {\em Progress in Quantum Electronics}, 4:35--110, 1975.

\bibitem{Sheik-Bahae1990a}
Mansoor Sheik-Bahae, A.A. Said, T.-H. Wei, D.J. Hagan, and E.W. {Van Stryland}.
\newblock {Sensitive measurement of optical nonlinearities using a single beam}.
\newblock {\em IEEE Journal of Quantum Electronics}, 26(4):760--769, apr 1990.

\bibitem{Chen2008}
L.~M. Chen, M.~Kando, M.~H. Xu, Y.~T. Li, J.~Koga, M.~Chen, H.~Xu, X.~H. Yuan, Q.~L. Dong, Z.~M. Sheng, S.~V. Bulanov, Y.~Kato, J.~Zhang, and T.~Tajima.
\newblock {Study of X-Ray Emission Enhancement via a High-Contrast Femtosecond Laser Interacting with a Solid Foil}.
\newblock {\em Physical Review Letters}, 100(4):045004, jan 2008.

\bibitem{Sun2022}
Ying Sun, Weiyi Yin, Qian Yao, Xiangyu Ren, Juan Song, and Ye~Dai.
\newblock {Temporal modulation toward femtosecond laser-induced nonlinear ionization process}.
\newblock {\em Optics Letters}, 47(23):6045, dec 2022.

\bibitem{Louzon2005}
E.~Louzon, Z.~Henis, S.~Pecker, Y.~Ehrlich, D.~Fisher, M.~Fraenkel, and A.~Zigler.
\newblock {Reduction of damage threshold in dielectric materials induced by negatively chirped laser pulses}.
\newblock {\em Applied Physics Letters}, 87(24):1--3, dec 2005.

\bibitem{Mareev2021}
Evgenii Mareev, Ekaterina Migal, and Fedor~V. Potemkin.
\newblock {The effect of chirp and wavelength for ultrafast bulk modification of solids with tightly focused laser pulses}.
\newblock In Vitaly~E. Gruzdev, Christopher~W. Carr, Detlev Ristau, and Carmen~S. Menoni, editors, {\em Laser-Induced Damage in Optical Materials 2021}, volume 11910, page~3. SPIE, nov 2021.

\bibitem{Duchateau2014}
Guillaume Duchateau and Antoine Bourgeade.
\newblock {Influence of the time-dependent pulse spectrum on ionization and laser propagation in nonlinear optical materials}.
\newblock {\em Physical Review A}, 89(5):053837, may 2014.

\bibitem{Bristow2007}
Alan~D. Bristow, Nir Rotenberg, and Henry~M. van Driel.
\newblock {Two-photon absorption and Kerr coefficients of silicon for 850–2200nm}.
\newblock {\em Applied Physics Letters}, 90(19):191104, may 2007.

\bibitem{Lin2007}
Q.~Lin, J.~Zhang, G.~Piredda, R.~W. Boyd, P.~M. Fauchet, and G.~P. Agrawal.
\newblock {Dispersion of silicon nonlinearities in the near infrared region}.
\newblock {\em Applied Physics Letters}, 91(2):021111, jul 2007.

\bibitem{Vignaud2004}
D.~Vignaud, J.~F. Lampin, and F.~Mollot.
\newblock {Two-photon absorption in InP substrates in the 1.55$\mu$m range}.
\newblock {\em Applied Physics Letters}, 85(2):239--241, jul 2004.

\bibitem{Hurlbut2007}
W.~C. Hurlbut, Yun-Shik Lee, K.~L. Vodopyanov, P.~S. Kuo, and M.~M. Fejer.
\newblock {Multiphoton absorption and nonlinear refraction of GaAs in the mid-infrared}.
\newblock {\em Optics Letters}, 32(6):668, mar 2007.

\bibitem{Gonzalez2009}
Leonel~P. Gonzalez, Joel~M. Murray, Srinivasan Krishnamurthy, and Shekhar Guha.
\newblock {Wavelength dependence of two photon and free carrier absorptions in InP}.
\newblock {\em Optics Express}, 17(11):8741, may 2009.

\bibitem{Krishnamurthy2011}
Srini Krishnamurthy, Zhi~Gang Yu, Leonel~P. Gonzalez, and Shekhar Guha.
\newblock {Temperature- and wavelength-dependent two-photon and free-carrier absorption in GaAs, InP, GaInAs, and InAsP}.
\newblock {\em Journal of Applied Physics}, 109(3), feb 2011.

\bibitem{Oishi2018}
Masaki Oishi, Tomohisa Shinozaki, Hikaru Hara, Kazunuki Yamamoto, Toshio Matsusue, and Hiroyuki Bando.
\newblock {Measurement of polarization dependence of two-photon absorption coefficient in InP using extended Z-scan technique for thick materials}.
\newblock {\em Japanese Journal of Applied Physics}, 57(3):030306, mar 2018.

\bibitem{Okhrimchuk2009}
A.~G. Okhrimchuk, V.~K. Mezentsev, H.~Schmitz, M.~Dubov, and I.~Bennion.
\newblock {Cascaded nonlinear absorption of femtosecond laser pulses in dielectrics}.
\newblock {\em Laser Physics}, 19(7):1415--1422, jul 2009.

\bibitem{Macdonald2010}
J.~R. Macdonald, R.~R. Thomson, S.~J. Beecher, N.~D. Psaila, H.~T. Bookey, and A.~K. Kar.
\newblock {Ultrafast laser inscription of near-infrared waveguides in polycrystalline ZnSe}.
\newblock {\em Optics Letters}, 35(23):4036, dec 2010.

\bibitem{Grivas2008}
C.~Grivas and R.~W. Eason.
\newblock {Fabrication of reflective volume gratings in pulsed-laser-deposited Ti:sapphire waveguides with UV femtosecond laser pulses}.
\newblock {\em Applied Physics A}, 93(1):219--223, oct 2008.

\bibitem{Saliminia2012}
Ali Saliminia, Jean-Philippe B{\'{e}}rub{\'{e}}, and R{\'{e}}al Vall{\'{e}}e.
\newblock {Refractive index-modified structures in glass written by 266nm fs laser pulses}.
\newblock {\em Optics Express}, 20(25):27410, dec 2012.

\bibitem{Baudisch2018}
Matthias Baudisch, Marcus Beutler, Martin Gebhardt, Christian Gaida, Fabian Stutzki, Steffen Hadrich, Robert Herda, Kevin Zawilski, Peter Schunemann, Armin Zach, Jens Limpert, and Ingo Rimke.
\newblock {2.3-12 $\mu$m tunable, sub-10 optical cycle, ZnGeP$_{2}$-based OPA directly pumped by a Tm:fiber laser at 1.96 $\mu$m and 100 kHz}.
\newblock In {\em Laser Congress 2018 (ASSL)}, volume Part F121-, page AW4A.3, Washington, D.C., 2018. OSA.

\bibitem{DeKernier2022}
Isaure {De Kernier}, Yann Wanwanscappel, David Boutolleau, Thomas Carmignani, Fabien Clop, Philippe Feautrier, Jean-Luc Gach, St{\'{e}}phane Lemarchand, and Eric Stadler.
\newblock {C-RED 2 ER: an extended range SWIR camera with applications in hyperspectral imaging}.
\newblock In {\em Proceedings of SPIE}, page 119970V. SPIE, mar 2022.

\bibitem{Li2021}
Qingfeng Li, Maxime Chambonneau, Markus Blothe, Herbert Gross, and Stefan Nolte.
\newblock {Flexible, fast, and benchmarked vectorial model for focused laser beams}.
\newblock {\em Applied Optics}, 60(13):3954, may 2021.

\bibitem{Li2021a}
Qingfeng Li.
\newblock {\href{https://github.com/QF06/InFocus}{InFocus} https://github.com/QF06/InFocus (accessed on April 2025)}.

\bibitem{Dimitrov1996}
Vesselin Dimitrov and Sumio Sakka.
\newblock {Linear and nonlinear optical properties of simple oxides. II}.
\newblock {\em Journal of Applied Physics}, 79(3):1741--1745, feb 1996.

\bibitem{Varshni1967}
Y.P. Varshni.
\newblock {Temperature dependence of the energy gap in semiconductors}.
\newblock {\em Physica}, 34(1):149--154, jan 1967.

\bibitem{Li1980}
H.~H. Li.
\newblock {Refractive index of silicon and germanium and its wavelength and temperature derivatives}.
\newblock {\em Journal of Physical and Chemical Reference Data}, 9(3):561--658, jul 1980.

\bibitem{DeLeonardis2016}
Francesco {De Leonardis}, Benedetto Troia, Richard~A. Soref, and Vittorio M.~N. Passaro.
\newblock {Dispersion of nonresonant third-order nonlinearities in GeSiSn ternary alloys}.
\newblock {\em Scientific Reports}, 6(1):32622, sep 2016.

\bibitem{Sohn2017}
B.-U. Sohn, C.~Monmeyran, L.~C. Kimerling, A.~M. Agarwal, and D.~T.~H. Tan.
\newblock {Kerr nonlinearity and multi-photon absorption in germanium at mid-infrared wavelengths}.
\newblock {\em Applied Physics Letters}, 111(9):091902, aug 2017.

\bibitem{Zlatanovic2010}
Sanja Zlatanovic, Jung~S. Park, Slaven Moro, Jose M.~Chavez Boggio, Ivan~B. Divliansky, Nikola Alic, Shayan Mookherjea, and Stojan Radic.
\newblock {Mid-infrared wavelength conversion in silicon waveguides using ultracompact telecom-band-derived pump source}.
\newblock {\em Nature Photonics}, 4(8):561--564, aug 2010.

\bibitem{Wang2013}
Ting Wang, Nalla Venkatram, Jacek Gosciniak, Yuanjing Cui, Guodong Qian, Wei Ji, and Dawn T.~H. Tan.
\newblock {Multi-photon absorption and third-order nonlinearity in silicon at mid-infrared wavelengths}.
\newblock {\em Optics Express}, 21(26):32192, dec 2013.

\bibitem{Zhang2014}
Lin Zhang, Anuradha~M. Agarwal, Lionel~C. Kimerling, and Jurgen Michel.
\newblock {Nonlinear Group IV photonics based on silicon and germanium: from near-infrared to mid-infrared}.
\newblock {\em Nanophotonics}, 3(4-5):247--268, aug 2014.

\bibitem{Ensley2019}
Trenton~R. Ensley and Neal~K. Bambha.
\newblock {Ultrafast nonlinear refraction measurements of infrared transmitting materials in the mid-wave infrared}.
\newblock {\em Optics Express}, 27(26):37940, dec 2019.

\bibitem{Jansonas2022}
Gaudenis Jansonas, Rimantas Budriūnas, Mikas Vengris, and Arūnas Varanavi{\v{c}}ius.
\newblock {Interferometric measurements of nonlinear refractive index in the infrared spectral range}.
\newblock {\em Optics Express}, 30(17):30507, aug 2022.

\bibitem{Pettit1965}
G.~D. Pettit and W.~J. Turner.
\newblock {Refractive Index of InP}.
\newblock {\em Journal of Applied Physics}, 36(6):2081--2081, jun 1965.

\bibitem{Ching1993}
W.~Y. Ching and Ming-Zhu Huang.
\newblock {Calculation of optical excitations in cubic semiconductors. III. Third-harmonic generation}.
\newblock {\em Physical Review B}, 47(15):9479--9491, apr 1993.

\bibitem{Yu2013}
Yi~Yu, Evarist Palushani, Mikkel Heuck, Nadezda Kuznetsova, Philip~Tr{\o}st Kristensen, Sara Ek, Dragana Vukovic, Christophe Peucheret, Leif~Katsuo Oxenl{\o}we, Sylvain Combri{\'{e}}, Alfredo de~Rossi, Kresten Yvind, and Jesper M{\o}rk.
\newblock {Switching characteristics of an InP photonic crystal nanocavity: Experiment and theory}.
\newblock {\em Optics Express}, 21(25):31047, dec 2013.

\bibitem{Heuck2013}
M.~Heuck, S.~Combri{\'{e}}, G.~Lehoucq, S.~Malaguti, G.~Bellanca, S.~Trillo, P.~T. Kristensen, J.~M{\o}rk, J.~P. Reithmaier, and A.~de~Rossi.
\newblock {Heterodyne pump probe measurements of nonlinear dynamics in an indium phosphide photonic crystal cavity}.
\newblock {\em Applied Physics Letters}, 103(18):181120, oct 2013.

\bibitem{Jiao2020}
Yuqing Jiao, Jos van~der Tol, Vadim Pogoretskii, Jorn van Engelen, Amir~Abbas Kashi, Sander Reniers, Yi~Wang, Xinran Zhao, Weiming Yao, Tianran Liu, Francesco Pagliano, Andrea Fiore, Xuebing Zhang, Zizheng Cao, Rakesh~Ranjan Kumar, Hon~Ki Tsang, Rene van Veldhoven, Tjibbe de~Vries, Erik-Jan Geluk, Jeroen Bolk, Huub Ambrosius, Meint Smit, and Kevin Williams.
\newblock {Indium Phosphide Membrane Nanophotonic Integrated Circuits on Silicon}.
\newblock {\em physica status solidi (a)}, 217(3):1900606, feb 2020.

\bibitem{Skauli2003}
T.~Skauli, P.~S. Kuo, K.~L. Vodopyanov, T.~J. Pinguet, O.~Levi, L.~A. Eyres, J.~S. Harris, M.~M. Fejer, B.~Gerard, L.~Becouarn, and E.~Lallier.
\newblock {Improved dispersion relations for GaAs and applications to nonlinear optics}.
\newblock {\em Journal of Applied Physics}, 94(10):6447--6455, nov 2003.

\bibitem{Dinu2003}
M.~Dinu, F.~Quochi, and H.~Garcia.
\newblock {Third-order nonlinearities in silicon at telecom wavelengths}.
\newblock {\em Applied Physics Letters}, 82(18):2954--2956, may 2003.

\bibitem{Zha2007}
Congji Zha, Rongping Wang, Anita Smith, Amrita Prasad, Ruth~A. Jarvis, and Barry Luther-Davies.
\newblock {Optical properties and structural correlations of GeAsSe chalcogenide glasses}.
\newblock {\em Journal of Materials Science: Materials in Electronics}, 18(S1):389--392, oct 2007.

\bibitem{Joerg2016}
Alexandre Jo{\"{e}}rg, Fabien Lemarchand, Mengxue Zhang, Michel Lequime, and Julien Lumeau.
\newblock {Optical characterization of photosensitive AMTIR-1 chalcogenide thin layers deposited by electron beam deposition}.
\newblock {\em Journal of Non-Crystalline Solids}, 442:22--28, 2016.

\bibitem{Miller1975}
A.~Miller and W.~Clark.
\newblock {Electrical Properties of ZnGeP$_{2}$ and CdGeP$_{2}$}.
\newblock {\em Le Journal de Physique Colloques}, 36(C3):C3--73--C3--75, sep 1975.

\bibitem{Boyd1971}
G.~D. Boyd, E.~Buehler, and F.~G. Storz.
\newblock {Linear and nonlinear optical properties of ZnGeP$_{2}$ and CdSe}.
\newblock {\em Applied Physics Letters}, 18(7):301--304, apr 1971.

\bibitem{Patel2020}
Abhishek Patel, Deobrat Singh, Yogesh Sonvane, P.B. Thakor, and Rajeev Ahuja.
\newblock {Bulk and monolayer As$_{2}$S$_{3}$ as promising thermoelectric material with high conversion performance}.
\newblock {\em Computational Materials Science}, 183(July):109913, oct 2020.

\bibitem{Rodney1958}
William~S. Rodney, Irving~H. Malitson, and Thomas~A. King.
\newblock {Refractive Index of Arsenic Trisulfide}.
\newblock {\em Journal of the Optical Society of America}, 48(9):633, sep 1958.

\bibitem{McCanny1977}
J.~V. McCanny and R.~B. Murray.
\newblock {The band structures of gallium and indium selenide}.
\newblock {\em Journal of Physics C: Solid State Physics}, 10(8):1211--1222, apr 1977.

\bibitem{Kato2013}
Kiyoshi Kato, Fumihito Tanno, and Nobuhiro Umemura.
\newblock {Sellmeier and thermo-optic dispersion formulas for GaSe (Revisited)}.
\newblock {\em Applied Optics}, 52(11):2325, apr 2013.

\bibitem{Kittel2005}
C.~Kittel.
\newblock {\em {Introduction to solid state physics}}.
\newblock John Wiley $\&$ Sons, Inc., 2005.

\bibitem{Bond1965}
W.~L. Bond.
\newblock {Measurement of the Refractive Indices of Several Crystals}.
\newblock {\em Journal of Applied Physics}, 36(5):1674--1677, may 1965.

\bibitem{Kawamori2022}
Taiki Kawamori, Peter~G. Schunemann, Vitaly Gruzdev, and Konstantin~L. Vodopyanov.
\newblock {High-order ( N = 4–6) multiphoton absorption and mid-infrared Kerr nonlinearity in GaP, ZnSe, GaSe, and ZGP crystals}.
\newblock {\em APL Photonics}, 7(8):086101, aug 2022.

\bibitem{Isik2020}
M.~Isik, H.H. Gullu, M.~Parlak, and N.M. Gasanly.
\newblock {Synthesis and temperature-tuned band gap characteristics of magnetron sputtered ZnTe thin films}.
\newblock {\em Physica B: Condensed Matter}, 582(November 2019):411968, apr 2020.

\bibitem{Li1984}
H.~H. Li.
\newblock {Refractive Index of ZnS, ZnSe, and ZnTe and Its Wavelength and Temperature Derivatives}.
\newblock {\em Journal of Physical and Chemical Reference Data}, 13(1):103--150, jan 1984.

\bibitem{Itoh1978}
Nobuo Itoh, Takemi Fujinaga, and Tanehiro Nakau.
\newblock {Birefringence in CdSiP$_2$}.
\newblock {\em Japanese Journal of Applied Physics}, 17(5):951--952, may 1978.

\bibitem{Wei2018}
Jean Wei, Joel~M. Murray, Frank~K. Hopkins, Douglas~M. Krein, Kevin~T. Zawilski, Peter~G. Schunemann, and Shekhar Guha.
\newblock {Measurement of refractive indices of CdSiP$_{2}$ at temperatures from 90 to 450 K}.
\newblock {\em Optical Materials Express}, 8(2):235, feb 2018.

\bibitem{Ferdinandus2020}
Manuel~R. Ferdinandus, Jamie~J. Gengler, Kent~L. Averett, Kevin~T. Zawilski, Peter~G. Schunemann, and Carl~M. Liebig.
\newblock {Nonlinear optical measurements of CdSiP$_{2}$ at near and mid-infrared wavelengths}.
\newblock {\em Optical Materials Express}, 10(9):2066, sep 2020.

\bibitem{Marcinkeviciute2018}
Agnė Marcinkevi{\v{c}}iūtė, Gintaras Tamo{\v{s}}auskas, and Audrius Dubietis.
\newblock {Supercontinuum generation in mixed thallous halides KRS-5 and KRS-6}.
\newblock {\em Optical Materials}, 78:339--344, apr 2018.

\bibitem{Rodney1956}
William~S. Rodney and Irving~H. Malitson.
\newblock {Refraction and Dispersion of Thallium Bromide Iodide}.
\newblock {\em Journal of the Optical Society of America}, 46(11):956, nov 1956.

\bibitem{Chen2007}
Shiyou Chen, X.~G. Gong, and Su-Huai Wei.
\newblock {Band-structure anomalies of the chalcopyrite semiconductors CuGaX$_{2}$ versus AgGaX$_{2}$ (X=S and Se) and their alloys}.
\newblock {\em Physical Review B}, 75(20):205209, may 2007.

\bibitem{Takaoka1999}
Eiko Takaoka and Kiyoshi Kato.
\newblock {Thermo-optic dispersion formula for AgGaS$_{2}$}.
\newblock {\em Applied Optics}, 38(21):4577, jul 1999.

\bibitem{Streetman2016}
S.~K. {Streetman, B. G., Banerjee}.
\newblock {\em {Solid State Electronic Devices}}.
\newblock Pearson Education Limited, 2016.

\bibitem{Marple1964}
D.~T.~F. Marple.
\newblock {Refractive Index of ZnSe, ZnTe, and CdTe}.
\newblock {\em Journal of Applied Physics}, 35(3):539--542, mar 1964.

\bibitem{Yelisseyev2020}
A.P. Yelisseyev, S.I. Lobanov, P.G. Krinitsin, and L.I. Isaenko.
\newblock {The optical properties of the nonlinear crystal BaGa$_{4}$Se$_{7}$}.
\newblock {\em Optical Materials}, 99(November 2019):109564, jan 2020.

\bibitem{Zhai2017}
Naixia Zhai, Chao Li, Bo~Xu, Lei Bai, Jiyong Yao, Guochun Zhang, Zhanggui Hu, and Yicheng Wu.
\newblock {Temperature-Dependent Sellmeier Equations of IR Nonlinear Optical Crystal BaGa$_{4}$Se$_{7}$}.
\newblock {\em Crystals}, 7(3):62, feb 2017.

\bibitem{Hettner1948}
G~Hettner and G~Leisegang.
\newblock {Die Dispersion der Mischkristalle TlBr-TlJ (KRS5) und TlCl-TlBr (KRS6) im Ultrarot}.
\newblock {\em Optik}, 3(4):305--314, 1948.

\bibitem{Klimm2014}
Detlef Klimm.
\newblock {Electronic materials with a wide band gap: recent developments}.
\newblock {\em IUCrJ}, 1(5):281--290, sep 2014.

\bibitem{Wang2013a}
Shunchong Wang, Minjie Zhan, Gang Wang, Hongwen Xuan, Wei Zhang, Chunjun Liu, Chunhua Xu, Yu~Liu, Zhiyi Wei, and Xiaolong Chen.
\newblock {4H-SiC: a new nonlinear material for midinfrared lasers}.
\newblock {\em Laser $\&$ Photonics Reviews}, 7(5):831--838, sep 2013.

\bibitem{Cardenas2015}
Jaime Cardenas, Mengjie Yu, Yoshitomo Okawachi, Carl~B. Poitras, Ryan K.~W. Lau, Avik Dutt, Alexander~L. Gaeta, and Michal Lipson.
\newblock {Optical nonlinearities in high-confinement silicon carbide waveguides}.
\newblock {\em Optics Letters}, 40(17):4138, sep 2015.

\bibitem{Barker1973}
A.~S. Barker and M.~Ilegems.
\newblock {Infrared Lattice Vibrations and Free-Electron Dispersion in GaN}.
\newblock {\em Physical Review B}, 7(2):743--750, jan 1973.

\bibitem{Almeida2019}
Gustavo F.~B. Almeida, Sabrina N.~C. Santos, Jonathas~P. Siqueira, Jessica Dipold, Tobias Voss, and Cleber~R. Mendon{\c{c}}a.
\newblock {Third-Order Nonlinear Spectrum of GaN under Femtosecond-Pulse Excitation from the Visible to the Near Infrared}.
\newblock {\em Photonics}, 6(2):69, jun 2019.

\bibitem{Debenham1984}
Mary Debenham.
\newblock {Refractive indices of zinc sulfide in the 0.405–13-$\mu$m wavelength range}.
\newblock {\em Applied Optics}, 23(14):2238, jul 1984.

\bibitem{Yelisseyev2012}
A.~Yelisseyev, Z.~S. Lin, M.~Starikova, L.~Isaenko, and S.~Lobanov.
\newblock {Optical transitions due to native defects in nonlinear optical crystals LiGaS$_{2}$}.
\newblock {\em Journal of Applied Physics}, 111(11):113507, jun 2012.

\bibitem{Vu2020}
Tuan~V. Vu, A.~A. Lavrentyev, B.~V. Gabrelian, Dat~D. Vo, Pham~D. Khang, L.~I. Isaenko, S.~I. Lobanov, A.~F. Kurus', and O.~Y. Khyzhun.
\newblock {Optical and electronic properties of lithium thiogallate (LiGaS$_{2}$): experiment and theory}.
\newblock {\em RSC Advances}, 10(45):26843--26852, 2020.

\bibitem{Singh2017}
Prashant Singh, Manoj~K. Harbola, and Duane~D. Johnson.
\newblock {Better band gaps for wide-gap semiconductors from a locally corrected exchange-correlation potential that nearly eliminates self-interaction errors}.
\newblock {\em Journal of Physics: Condensed Matter}, 29(42):424001, oct 2017.

\bibitem{Luke2015}
Kevin Luke, Yoshitomo Okawachi, Michael R.~E. Lamont, Alexander~L. Gaeta, and Michal Lipson.
\newblock {Broadband mid-infrared frequency comb generation in a Si$_{3}$N$_{4}$ microresonator}.
\newblock {\em Optics Letters}, 40(21):4823, nov 2015.

\bibitem{Ikeda2008}
Kazuhiro Ikeda, Robert~E. Saperstein, Nikola Alic, and Yeshaiahu Fainman.
\newblock {Thermal and Kerr nonlinear properties of plasma-deposited silicon nitride/ silicon dioxide waveguides}.
\newblock {\em Optics Express}, 16(17):12987, aug 2008.

\bibitem{Neufeld2022}
S.~Neufeld, Arno Schindlmayr, and W.~G. Schmidt.
\newblock {Quasiparticle energies and optical response of RbTiOPO$_{4}$ and KTiOAsO$_{4}$}.
\newblock {\em Journal of Physics: Materials}, 5(1):015002, jan 2022.

\bibitem{Cheng1993}
L.~K. Cheng, L.-T. Cheng, J.~D. Bierlein, F.~C. Zumsteg, and A.~A. Ballman.
\newblock {Properties of doped and undoped crystals of single domain KTiOAsO$_{4}$}.
\newblock {\em Applied Physics Letters}, 62(4):346--348, jan 1993.

\bibitem{Phillip1964}
H.~R. Phillip and E.~A. Taft.
\newblock {Kramers-Kronig Analysis of Reflectance Data for Diamond}.
\newblock {\em Physical Review}, 136(5A):A1445--A1448, nov 1964.

\bibitem{Hausmann2014}
B.~J.~M. Hausmann, I.~Bulu, V.~Venkataraman, P.~Deotare, and M.~Lon{\v{c}}ar.
\newblock {Diamond nonlinear photonics}.
\newblock {\em Nature Photonics}, 8(5):369--374, may 2014.

\bibitem{Pastrnak1966}
J.~Pastrň{\'{a}}k and L.~Roskovcov{\'{a}}.
\newblock {Refraction Index Measurements on AlN Single Crystals}.
\newblock {\em physica status solidi (b)}, 14(1):K5--K8, 1966.

\bibitem{Jung2013}
Hojoong Jung, Chi Xiong, King~Y. Fong, Xufeng Zhang, and Hong~X. Tang.
\newblock {Optical frequency comb generation from aluminum nitride microring resonator}.
\newblock {\em Optics Letters}, 38(15):2810, aug 2013.

\bibitem{Malitson1962}
Irving~H. Malitson.
\newblock {Refraction and Dispersion of Synthetic Sapphire}.
\newblock {\em Journal of the Optical Society of America}, 52(12):1377, dec 1962.

\bibitem{Patwardhan2021}
Gauri~N. Patwardhan, Jared~S. Ginsberg, Cecilia~Y. Chen, M.~Mehdi Jadidi, and Alexander~L. Gaeta.
\newblock {Nonlinear refractive index of solids in mid-infrared}.
\newblock {\em Optics Letters}, 46(8):1824, apr 2021.

\bibitem{Fischetti1985}
M.~V. Fischetti, D.~J. DiMaria, S.~D. Brorson, T.~N. Theis, and J.~R. Kirtley.
\newblock {Theory of high-field electron transport in silicon dioxide}.
\newblock {\em Physical Review B}, 31(12):8124--8142, jun 1985.

\bibitem{Malitson1965}
I.~H. Malitson.
\newblock {Interspecimen Comparison of the Refractive Index of Fused Silica}.
\newblock {\em Journal of the Optical Society of America}, 55(10):1205, oct 1965.

\bibitem{Tsujibayashi2002}
Toru Tsujibayashi, Koichi Toyoda, Shiro Sakuragi, Masao Kamada, and Minoru Itoh.
\newblock {Spectral profile of the two-photon absorption coefficients in CaF$_{2}$ and BaF$_{2}$}.
\newblock {\em Applied Physics Letters}, 80(16):2883--2885, apr 2002.

\bibitem{Li1980a}
H.~H. Li.
\newblock {Refractive index of alkaline earth halides and its wavelength and temperature derivatives}.
\newblock {\em Journal of Physical and Chemical Reference Data}, 9(1):161--290, jan 1980.

\bibitem{Chaney1971}
Roy~C. Chaney, Earl~E. Lafon, and Chun~C. Lin.
\newblock {Energy Band Structure of Lithium Fluoride Crystals by the Method of Tight Binding}.
\newblock {\em Physical Review B}, 4(8):2734--2741, oct 1971.

\bibitem{Li1976}
H.~H. Li.
\newblock {Refractive index of alkali halides and its wavelength and temperature derivatives}.
\newblock {\em Journal of Physical and Chemical Reference Data}, 5(2):329--528, apr 1976.

\bibitem{Thomas1973}
J.~Thomas, G.~Stephan, J.~C. Lemonnier, M.~Nisar, and S.~Robin.
\newblock {Optical Anisotropy of MgF, in Its UV Absorption Region}.
\newblock {\em Physica Status Solidi (b)}, 56(1):163--170, mar 1973.

\bibitem{Fibich2000}
Gadi Fibich and Alexander~L. Gaeta.
\newblock {Critical power for self-focusing in bulk media and in hollow waveguides}.
\newblock {\em Optics Letters}, 25(5):335, 2000.

\bibitem{Hon2011}
Nick~K. Hon, Richard Soref, and Bahram Jalali.
\newblock {The third-order nonlinear optical coefficients of Si, Ge, and Si$_{1-x}$Ge$_{x}$ in the midwave and longwave infrared}.
\newblock {\em Journal of Applied Physics}, 110(1):011301, jul 2011.

\bibitem{Garcia2012}
Hernando Garcia and Kobra~Nasiri Avanaki.
\newblock {Direct and indirect two-photon absorption in Ge within the effective mass approximation}.
\newblock {\em Applied Physics Letters}, 100(13):131105, mar 2012.

\bibitem{Mashanovich2017}
Goran~Z Mashanovich, Colin~J Mitchell, Jordi~Soler Penades, Ali~Z Khokhar, Callum~G Littlejohns, Wei Cao, Zhibo Qu, Stevan Stankovic, Frederic~Y Gardes, Taha~Ben Masaud, Harold M~H Chong, Vinita Mittal, Ganapathy~Senthil Murugan, James~S Wilkinson, Anna~C Peacock, and Milos Nedeljkovic.
\newblock {Germanium Mid-Infrared Photonic Devices}.
\newblock {\em Journal of Lightwave Technology}, 35(4):624--630, feb 2017.

\bibitem{Euser2005}
Tijmen~G. Euser and Willem~L. Vos.
\newblock {Spatial homogeneity of optically switched semiconductor photonic crystals and of bulk semiconductors}.
\newblock {\em Journal of Applied Physics}, 97(4):043102, feb 2005.

\bibitem{Tiedje2007}
H.F. Tiedje, H.K. Haugen, and J.S. Preston.
\newblock {Measurement of nonlinear absorption coefficients in GaAs, InP and Si by an optical pump THz probe technique}.
\newblock {\em Optics Communications}, 274(1):187--197, jun 2007.

\bibitem{Benis2020}
Sepehr Benis, Claudiu~M. Cirloganu, Nicholas Cox, Trenton Ensley, Honghua Hu, Gero Nootz, Peter~D. Olszak, Lazaro~A. Padilha, Davorin Peceli, Matthew Reichert, Scott Webster, Milton Woodall, David~J. Hagan, and Eric~W. {Van Stryland}.
\newblock {Three-photon absorption spectra and bandgap scaling in direct-gap semiconductors}.
\newblock {\em Optica}, 7(8):888, aug 2020.

\bibitem{Peceli2013}
Davorin Peceli, Peter~D. Olszak, Claudiu~M. Cirloganu, Scott Webster, Lazaro~A. Padilha, Trenton Ensley, Honghua Hu, Gero Nootz, David~J. Hagan, and Eric~W. {Van Stryland}.
\newblock {Three-Photon Absorption of GaAs and other Semiconductors}.
\newblock In {\em Nonlinear Optics}, page NTu1B.6, Washington, D.C., 2013. OSA.

\bibitem{Bourgeade2015}
A~Bourgeade, T~Donval, L~Gallais, L~Lamaign{\`{e}}re, and J.-L. Rullier.
\newblock {Modeling surface defects in fused silica optics for laser wave propagation}.
\newblock {\em Journal of the Optical Society of America B}, 32(4):655, apr 2015.

\bibitem{Chambonneau2015b}
Maxime Chambonneau, Margaux Chanal, St{\'{e}}phane Reyn{\'{e}}, Guillaume Duchateau, Jean-yves Natoli, and Laurent Lamaign{\`{e}}re.
\newblock {Investigations on laser damage growth in fused silica with simultaneous wavelength irradiation}.
\newblock {\em Applied Optics}, 54(6):1463, feb 2015.

\bibitem{Chambonneau2021b}
M.~Chambonneau, M.~Blothe, Q.~Li, V.~Yu. Fedorov, T.~Heuermann, M.~Gebhardt, C.~Gaida, S.~Tertelmann, F.~Sotier, J.~Limpert, S.~Tzortzakis, and S.~Nolte.
\newblock {Transverse ultrafast laser inscription in bulk silicon}.
\newblock {\em Physical Review Research}, 3(4):043037, oct 2021.

\bibitem{Salzberg1957}
Calvin~D. Salzberg and John~J. Villa.
\newblock {Infrared Refractive Indexes of Silicon Germanium and Modified Selenium Glass}.
\newblock {\em Journal of the Optical Society of America}, 47(3):244, mar 1957.

\bibitem{Burnett2016}
John~H. Burnett, Simon~G. Kaplan, Eric Stover, and Adam Phenis.
\newblock {Refractive index measurements of Ge}.
\newblock In Paul~D. LeVan, Ashok~K. Sood, Priyalal Wijewarnasuriya, and Arvind~I. D'Souza, editors, {\em Infrared Sensors, Devices, and Applications VI}, volume 9974, page 99740X, sep 2016.

\end{thebibliography}
}


\pagebreak
\begin{center}
\twocolumn[\begin{@twocolumnfalse}
\textbf{\LARGE \textcolor{blue}{Supplementary Information: The universality of filamentation-caused challenges of ultrafast laser energy deposition in semiconductors} \vspace{30px}}
\end{@twocolumnfalse}]
\end{center}
\setcounter{equation}{0}
\setcounter{figure}{0}
\setcounter{table}{0}
\setcounter{page}{1}
\setcounter{section}{0}
\makeatletter
\renewcommand{\theequation}{S\arabic{equation}}
\renewcommand{\thesection}{S\arabic{section}}
\renewcommand{\thesubsection}{S\arabic{section}.\arabic{subsection}}
\renewcommand{\thefigure}{S\arabic{figure}}
\renewcommand{\thetable}{S\arabic{table}}

\captionsetup{format=plain, font=small, labelfont=bf}

\tableofcontents

\section{Optical properties}
\label{sec:OptParam}

As a state of the art for linear and nonlinear refraction in various solids, the band gap $E_{g}$, the linear and nonlinear refractive indices ($n_{0}$ and $n_{2}$, respectively) and the corresponding references are given in Table~\ref{tab:OptParam} for a selection of 30 materials. These data are displayed in the Primary Manuscript, Fig.~\ref{fig:Fig1}, where the relations $1-(n_{0}^{2}-1)/(n_{0}^{2}+2)=\sqrt{E_{g}/20}$ \cite{Dimitrov1996}, and $n_{2}\propto E_{g}^{-3}$ are verified (blue and red dashed curves, respectively).\\

\begin{table}[!ht]
  \centering
  \caption{Band gap $E_{g}$, linear, and nonlinear refractive indices ($n_{0}$ and $n_{2}$, respectively) of 30 materials. The data are given at 300~K, and at a wavelength of $\lambda=1960$~nm or close to. For birefringent materials, $n_{0}$ is averaged over the ordinary and extraordinary refractive indices.\\}
  \label{tab:OptParam}
  \scalebox{0.58}{\begin{tabular}{@{}ccccccc@{}}
    \toprule
    Material & $E_{g}$ (eV) &  & $n_{0}$ &  & $n_{2}$ (m$^{2}$/W) & \\
    \midrule
    Ge & 0.66 & \cite{Varshni1967} & 4.11 & \cite{Li1980} &  $1.20\times10^{-17}$ & \cite{DeLeonardis2016,Sohn2017} \\
    Si & 1.11 & \cite{Varshni1967} & 3.45 & \cite{Li1980} & $8.71\times10^{-18}$ & \cite{Bristow2007,Lin2007,Zlatanovic2010,Wang2013,Zhang2014,Ensley2019,Jansonas2022} \\
    InP & 1.27 & \cite{Varshni1967} & 3.13 & \cite{Pettit1965} & $2.57\times10^{-17}$ & \cite{Ching1993,Yu2013,Heuck2013,Ensley2019,Jiao2020} \\
    GaAs & 1.43 & \cite{Varshni1967} & 3.34 & \cite{Skauli2003} &  $1.31\times10^{-17}$ & \cite{Dinu2003,Hurlbut2007,Ensley2019,Jansonas2022} \\
    AMTIR-1 (Ge$_{33}$As$_{12}$Se$_{55}$) & 1.81 & \cite{Zha2007} & 2.68 & \cite{Joerg2016} & $3.38\times10^{-18}$ & \cite{Ensley2019} \\
    ZGP (ZnGeP$_{2}$) & 2.00 & \cite{Miller1975} & 3.17 & \cite{Boyd1971} & $6.55\times10^{-18}$ & \cite{Jansonas2022} \\
    AMTIR-6 (As$_{2}$S$_{3}$) & 2.08 & \cite{Patel2020} & 2.43 & \cite{Rodney1958} & $2.50\times10^{-18}$ & \cite{Ensley2019} \\
    GaSe & 2.10 & \cite{McCanny1977} & 2.58 & \cite{Kato2013} & $2.14\times10^{-18}$ & \cite{Jansonas2022} \\
    GaP & 2.25 & \cite{Kittel2005} & 3.04 & \cite{Bond1965} & $1.80\times10^{-18}$ & \cite{Kawamori2022} \\
    ZnTe & 2.26 & \cite{Isik2020} & 2.72 & \cite{Li1984} & $3.10\times10^{-18}$ & \cite{Jansonas2022} \\
    CdSiP$_{2}$ & 2.45 & \cite{Itoh1978} & 3.06 & \cite{Wei2018} & $1.00\times10^{-18}$ & \cite{Ferdinandus2020} \\
    KRS-5 (TlBr-TlI) & 2.50 & \cite{Marcinkeviciute2018} & 2.40 & \cite{Rodney1956} & $2.60\times10^{-18}$ & \cite{Jansonas2022} \\
    AGS (AgGaS$_{2}$) & 2.65 & \cite{Chen2007} & 2.39 & \cite{Takaoka1999} & $2.31\times10^{-18}$ & \cite{Jansonas2022} \\
    ZnSe & 2.70 & \cite{Streetman2016} & 2.44 & \cite{Marple1964} & $1.50\times10^{-18}$ & \cite{Ensley2019} \\
    BGSe (BaGa$_{4}$Se$_{7}$) & 2.73 & \cite{Yelisseyev2020} & 2.46 & \cite{Zhai2017} & $1.96\times10^{-18}$ & \cite{Jansonas2022} \\
    KRS-6 (TlBr-TlCl) & 3.25 & \cite{Marcinkeviciute2018} & 2.21 & \cite{Hettner1948} & $1.21\times10^{-18}$ & \cite{Jansonas2022} \\
    4H-SiC & 3.27 & \cite{Klimm2014} & 2.58 & \cite{Wang2013a} & $8.60\times10^{-19}$ & \cite{Cardenas2015} \\
    GaN & 3.40 & \cite{Streetman2016} & 2.30 & \cite{Barker1973} & $9.02\times10^{-19}$ & \cite{Almeida2019} \\
    ZnS & 3.60 & \cite{Streetman2016} & 2.27 & \cite{Debenham1984} & $5.50\times10^{-19}$ & \cite{Ensley2019} \\
    LGS (LiGaS$_{2}$) & 3.93 & \cite{Yelisseyev2012} & 2.05 & \cite{Vu2020} & $5.70\times10^{-19}$ & \cite{Jansonas2022} \\
    Si$_{3}$N$_{4}$ & 5.20 & \cite{Singh2017} & 1.98 & \cite{Luke2015} & $2.40\times10^{-19}$ & \cite{Ikeda2008} \\
    KTA (KTiOAsO$_{4}$) & 5.24 & \cite{Neufeld2022} & 1.79 & \cite{Cheng1993} & $1.75\times10^{-19}$ & \cite{Jansonas2022} \\
    Diamond & 5.48 & \cite{Klimm2014} & 2.38 & \cite{Phillip1964} & $8.20\times10^{-20}$ & \cite{Hausmann2014} \\
    AlN & 6.20 & \cite{Klimm2014} & 2.13 & \cite{Pastrnak1966} & $2.30\times10^{-19}$ & \cite{Jung2013} \\
    Sapphire (Al$_{2}$O$_{3}$) & 8.30 & \cite{Klimm2014} & 1.74 & \cite{Malitson1962} & $2.44\times10^{-20}$ & \cite{Patwardhan2021} \\
    Fused silica (a-SiO$_{2}$) & 9.00 & \cite{Fischetti1985} & 1.44 & \cite{Malitson1965} & $2.80\times10^{-20}$ & \cite{Ensley2019} \\
    BaF$_{2}$ & 10.6 & \cite{Tsujibayashi2002} & 1.46 & \cite{Li1980a} & $3.20\times10^{-20}$ & \cite{Ensley2019} \\
    LiF & 10.9 & \cite{Chaney1971} & 1.38 & \cite{Li1976} & $9.80\times10^{-21}$ & \cite{Ensley2019} \\
    CaF$_{2}$ & 11.8 & \cite{Tsujibayashi2002} & 1.42 & \cite{Li1980a} & $1.70\times10^{-20}$ & \cite{Ensley2019} \\
    MgF$_{2}$ & 12.4 & \cite{Thomas1973} & 1.37 & \cite{Li1980a} & $1.10\times10^{-20}$ & \cite{Ensley2019} \\
    \bottomrule
  \end{tabular}}
\end{table}

Concentrating more specifically on the semiconductors investigated in this study, the nonlinear refraction and absorption coefficients are summarized in Table~\ref{tab:materials}.

\begin{table}[!ht]
  \centering
  \caption{Nonlinear refraction and absorption coefficients for the investigated semiconductors at a wavelength of $\lambda=1960$~nm or close to. Here, $P_{\rm{cr}}$, $\beta_{2}$, and $\beta_{3}$ are the critical power for self-focusing, the 2- and 3-photon absorption coefficient, respectively. The critical power is calculated as $P_{\rm{cr}}=\alpha \lambda^{2}/(4\pi n_{0} n_{2})$, where $\alpha = 1.8962$ for Gaussian beams \cite{Fibich2000}, and $n_{0}$ as well as $n_{2}$ are given in Table~\ref{tab:OptParam}.}
  \label{tab:materials}
  \scalebox{0.75}{\begin{tabular}{@{}ccccc@{}}
    \toprule
    \textbf{Material} & $P_{\rm{cr}}$ (kW) & $\beta_{2}$ (m/W) & $\beta_{3}$ (m$^{3}$/W$^{2}$) & Reference\\
    \midrule
    Ge & 12.2 & $7.10 \times 10^{-9}$ & -- & \cite{Hon2011,Garcia2012,Mashanovich2017,Sohn2017}\\
    Si & 20.1 & $1.96 \times 10^{-12}$ & -- & \cite{Bristow2007,Lin2007,Euser2005}\\
    InP & 7.5 & -- & $1.41 \times 10^{-25}$ & \cite{Tiedje2007,Benis2020}\\
    GaAs & 13.8 & -- & $3.84 \times 10^{-25}$ & \cite{Wherrett1984,Tiedje2007,Hurlbut2007,Peceli2013,Benis2020}\\
    \bottomrule
  \end{tabular}}
\end{table}

\section{Metrology}
\label{sec:metrology}
\subsection{Necessary imaging conditions}

Three conditions must be fulfilled for reliable nonlinear propagation imaging with high dynamic range. First, the numerical aperture of the imaging lens must be higher than the one of the focusing lens so that all angular components are imaged. This is guaranteed by the design of the optical arrangement for nonlinear propagation imaging, as we employed objective lenses with $\rm{NA}=0.40$ for focusing, and $\rm{NA}=0.85$ for imaging.\\

Second, for improved imaging performance, the camera response must be linear at the employed wavelength. To examine this, let us consider a laser pulse with a Gaussian temporal intensity profile $I(t)=I_{0}\exp(-4\ln(2)t^{2}/\tau^{2})$, where $\tau$ is the pulse duration defined at full width at half maximum (FWHM), $I_{0} \propto E/(\pi w_{0}^2 \tau)$ is the maximum intensity, $E$ is the pulse energy, and $w_{0}$ is the beam radius at $1/e^{2}$. The pixel amplitude $A$ on the camera is proportional to the free-electron density $n_{e}$ generated on the camera chip. In the general case of $N$-photon absorption, the following rate equation can be used to determine $n_{e}$
\begin{equation}
  \frac{\partial n_{e}}{\partial t}=\frac{\beta_{N}I^{N}}{N \hbar \omega},
\end{equation}
where $\beta_{N}$ is the $N$-photon absorption coefficient, $\hbar$ is the reduced Planck constant, and $\omega$ is the angular frequency of the illuminating laser. The total free-electron density produced by the pulse thus reads
\begin{equation}
  n_{e} \propto \left( \frac{E}{\pi w_{0}^2 \tau} \right)^{N} \int_{-\infty}^{+\infty} \exp{\left( -4N\ln{2}\frac{t^{2}}{\tau^{2}} \right)} dt,
\end{equation}
which leads to
\begin{equation}
\label{eq:camresp}
  n_{e} \propto \frac{E^{N}}{w_{0}^{2N} \tau^{N-1}}.
\end{equation}

\begin{figure}
\centering
\includegraphics[width=\linewidth]{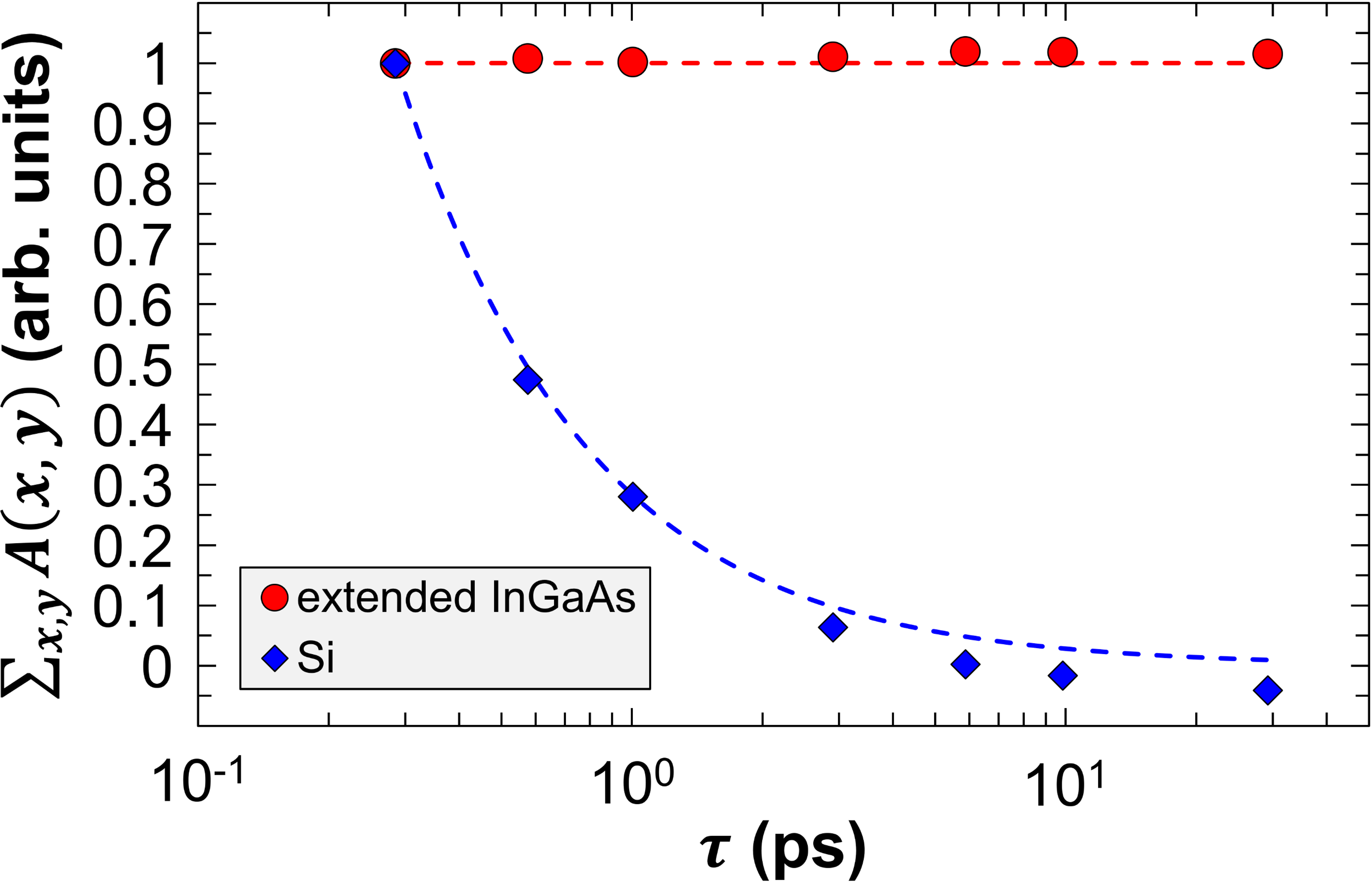}
\caption{\label{fig:FigSLinearity} Evolution of the sum of the pixel amplitudes $\sum_{x,y} A(x,y)$ as a function of the pulse duration $\tau$ for an extended InGaAs (red) and a Si camera (blue). The points are measurements in the linear propagation regime at a wavelength of 1960~nm, and the curves correspond to trends expected from Eq.~\eqref{eq:camresp}. The pulse energy $E$ and beam radius $w_{0}$ are kept constant in all measurements.}
\end{figure}

From Eq.~\eqref{eq:camresp}, one can note that the pixel amplitude $A$ is independent of the pulse duration $\tau$ for linear absorption ($N=1$), while it scales as $1/\tau$ for 2-photon absorption ($N=2$). This provides a method to verify experimentally the linearity of the camera response. This method consists of measuring the evolution of the sum of the pixel amplitudes $\sum_{x,y} A(x,y)$ as a function of $\tau$ while all other parameters---in particular $E$ and $w_{0}$---are kept constant. Experimental measurements at a wavelength of $\lambda=1960$~nm are displayed in Fig.~\ref{fig:FigSLinearity} for the extended InGaAs array used in our study (in red), and for a Si-based camera (in blue). As expected, the signal detected on the Si camera decreases with $\tau$. The deviation from the $1/\tau$ trend for the longest durations originates from the low signal-to-noise ratio. In contrast, the signal on the extended InGaAs array is independent of $\tau$, thus showing that absorption at $\lambda=1960$~nm on the camera chip is linear. While nonlinearly responding cameras ($N \ge 2$) can in principle be utilized for nonlinear propagation imaging to determine the maximum fluence \cite{Mareev2020}, this type of device underperforms linearly responding cameras ($N=1$) for observing fine features in the 3D fluence distribution, which requires a high dynamic range.\\

\begin{figure}
\centering
\includegraphics[width=\linewidth]{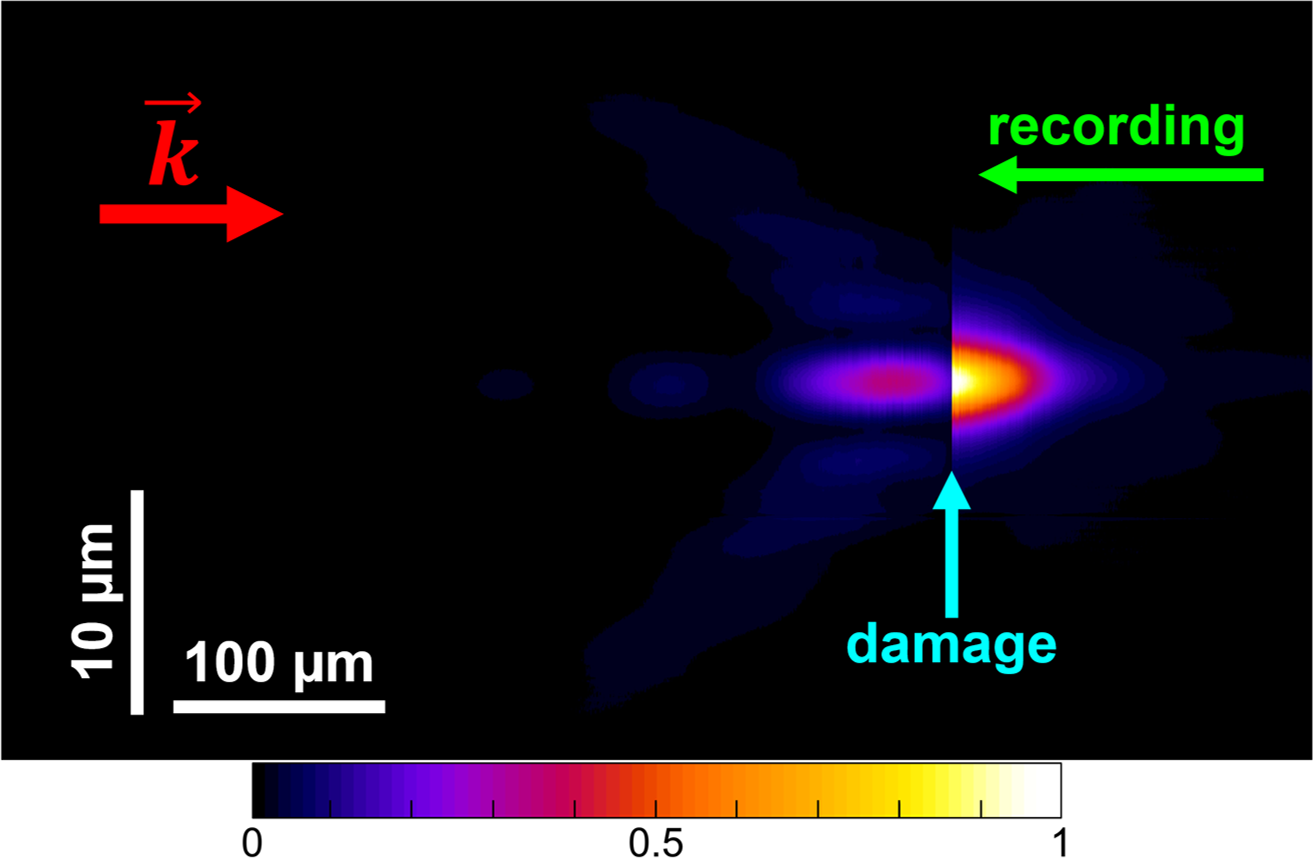}
\caption{\label{fig:FigSdamage} Normalized fluence distribution in the case of damage formed during nonlinear propagation imaging in InP for $E_{\rm{in}}=100$~nJ and $\tau=3$~ps. The vector $\vec{k}$ indicates the direction of laser propagation. The green and blue arrows indicate the recording chronology and the plane where damage has been formed, respectively.}
\end{figure}

The third condition to be fulfilled is that damage must not be induced in the sample during nonlinear propagation imaging. This has been systematically inspected after recording under white light illumination with the same microscope as for the measurements. When damage forms during the imaging procedure, it strongly affects the transmission by scattering and absorbing light. Moreover, during the recording of the subsequent images, the propagation can be strongly affected \cite{Bourgeade2015}. The fluence distributions obtained are all the more complex to interpret when one keeps in mind that damage formed on the surface or in the bulk of the sample can grow on a pulse-to-pulse basis \cite{Chambonneau2015b,Chambonneau2021b}. All these effects lead to the conclusion that damage must absolutely be avoided to get reliable experimental data, as illustrated in Fig.~\ref{fig:FigSdamage} where damage formed during the imaging procedure drastically modifies light transmission for the subsequently recorded images.

\subsection{Absence of cumulative effects}

\begin{figure}[!ht]
\centering
\includegraphics[width=\linewidth]{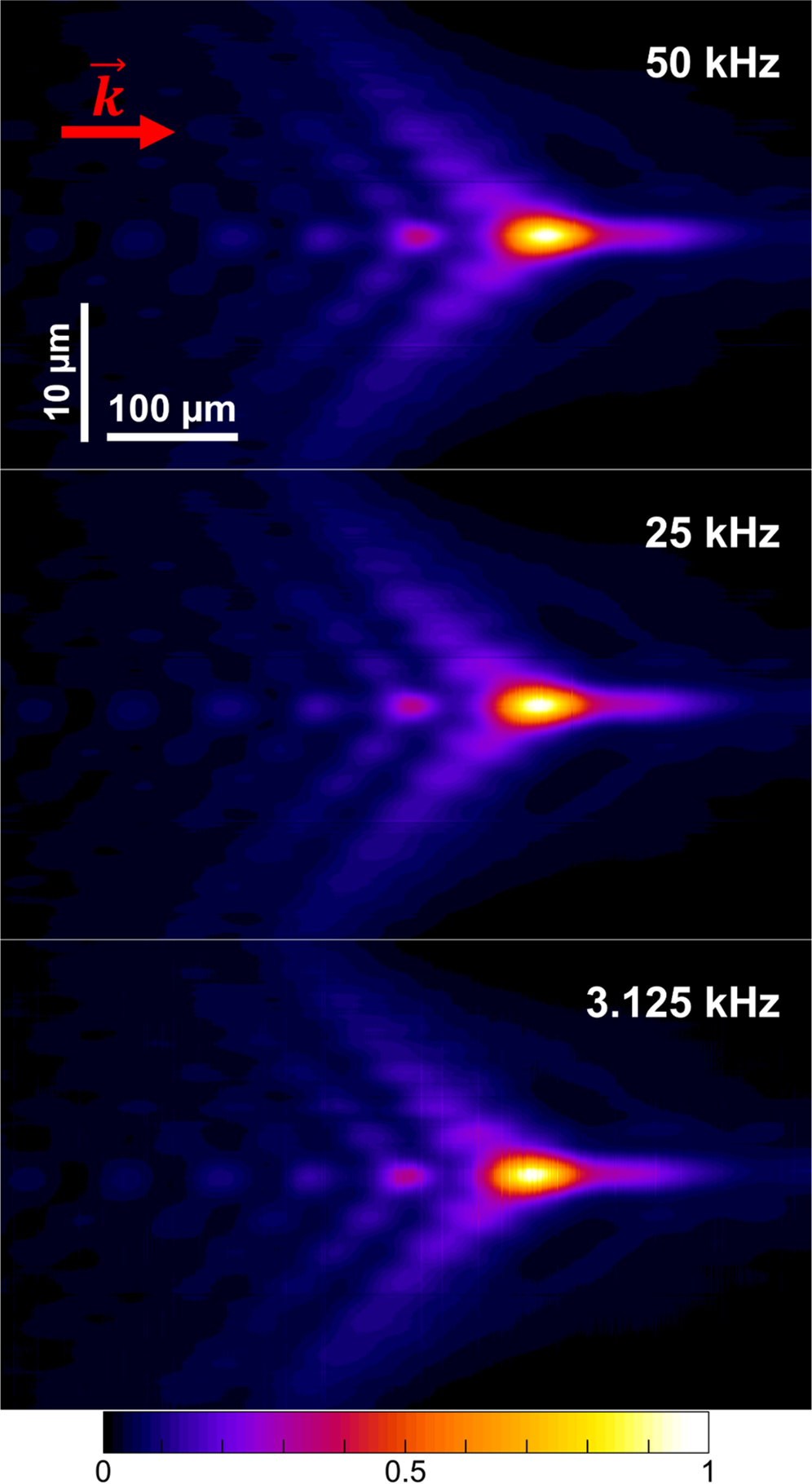}
\caption{\label{fig:FigSnoaccumulation} Normalized fluence distributions obtained at different repetition rates in Si for $E_{\rm{in}}=1000$~nJ and $\tau=275$~fs. The vector $\vec{k}$ indicates the direction of laser propagation. The spatial scales apply to all images.}
\end{figure}

In our nonlinear propagation measurements, we implicitly assumed that the material excited by one pulse returns to rest before the subsequent pulse arrives, i.e., there are no cumulative effects. Let us examine the validity of this assumption. In dielectrics, the thermal diffusivity is two orders of magnitude higher than in semiconductors. Consequently, extreme repetition rates---far above the repetition rate $\Omega=50$~kHz employed in our nonlinear propagation imaging measurements---are required to cause cumulative effects in these materials \cite{Wang2020a}. To verify this, we have performed nonlinear propagation imaging for different repetition rates, adjusted with an acousto-optic modulator implemented in the laser. As shown in Fig.~\ref{fig:FigSnoaccumulation}, for $\Omega \le 50$~kHz, the measured fluence distributions obtained in Si for high-intensity pulses ($E_{\rm{in}}=1000$~nJ and $\tau=275$~fs) are independent of the repetition rate. We thus conclude that no cumulative effects take place during the nonlinear propagation measurements.

\subsection{Repeatability}

\begin{figure}
\centering
\includegraphics[width=\linewidth]{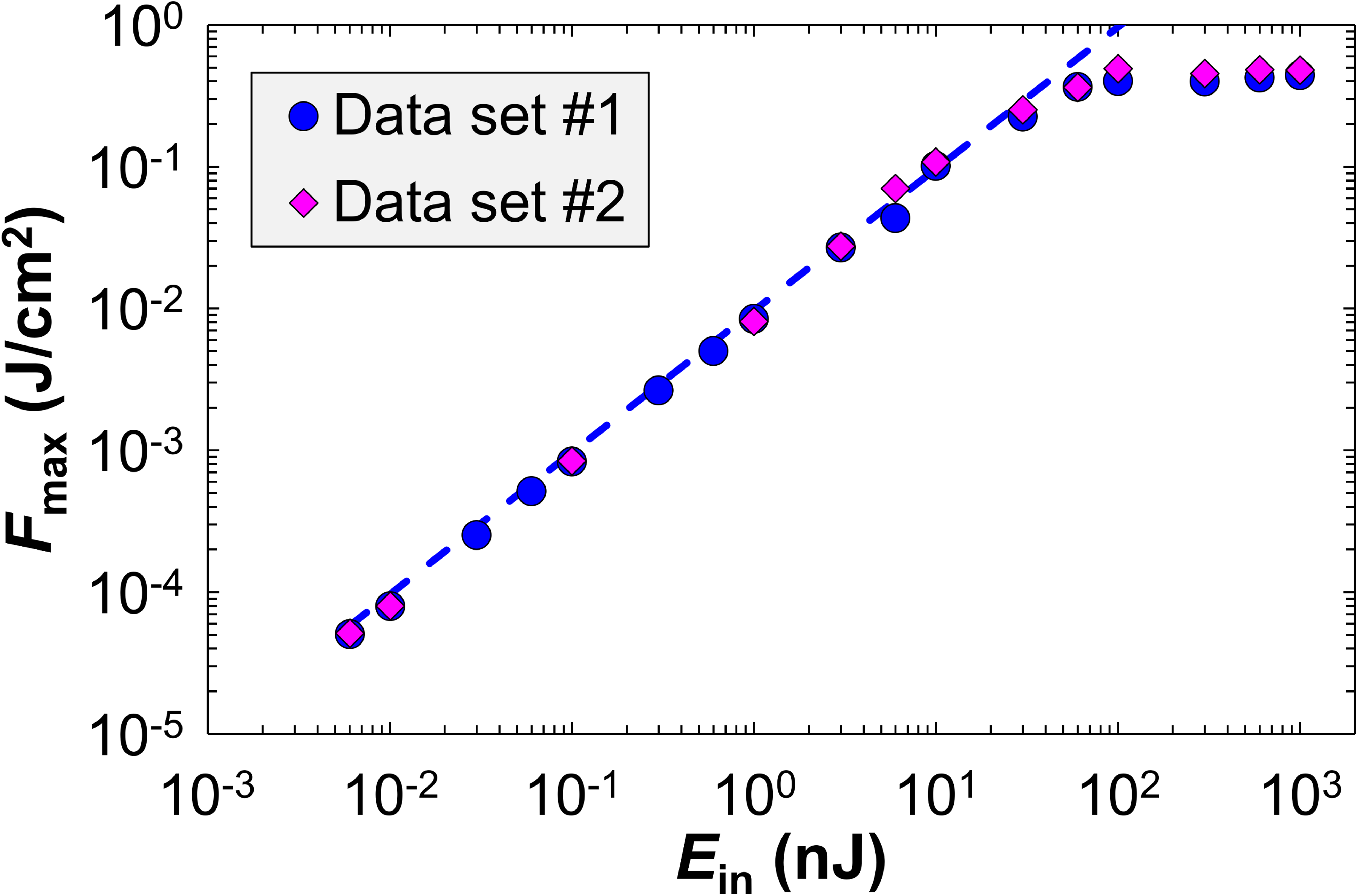}
\caption{\label{fig:FigSrepeatability} Evolution of the maximum fluence $F_{\rm{max}}$ measured in the bulk of Si on different days as a function of the input pulse energy $E_{\rm{in}}$. The pulse duration is $\tau=3$~ps. The dashed blue line corresponds to the linear regime.}
\end{figure}

To estimate the repeatability of our experimental method, nonlinear propagation imaging has been carried out on two different days in identical conditions (Si, $\tau=3$~ps). As shown in Fig.~\ref{fig:FigSrepeatability}, the measured $F_{\rm{max}}$ are similar over a broad range of input pulse energies $E_{\rm{in}}$. The input pulse energy value for which the experimental data deviate from the linear propagation regime is the same for both data sets ($E_{\rm{in}} = 30$~nJ). For $E_{\rm{in}} \geq 100$~nJ where the saturation plateau for $F_{\rm{max}}$ is reached, the peak fluence is similar for both measurements ($F_{p} \approx 0.31$~J/cm$^{2}$), with a standard deviation of $15\%$.

\section{Linear propagation regime}
\label{sec:linpropregime}

\subsection{Linear propagation calculations}

\begin{figure}
\centering
\includegraphics[width=\linewidth]{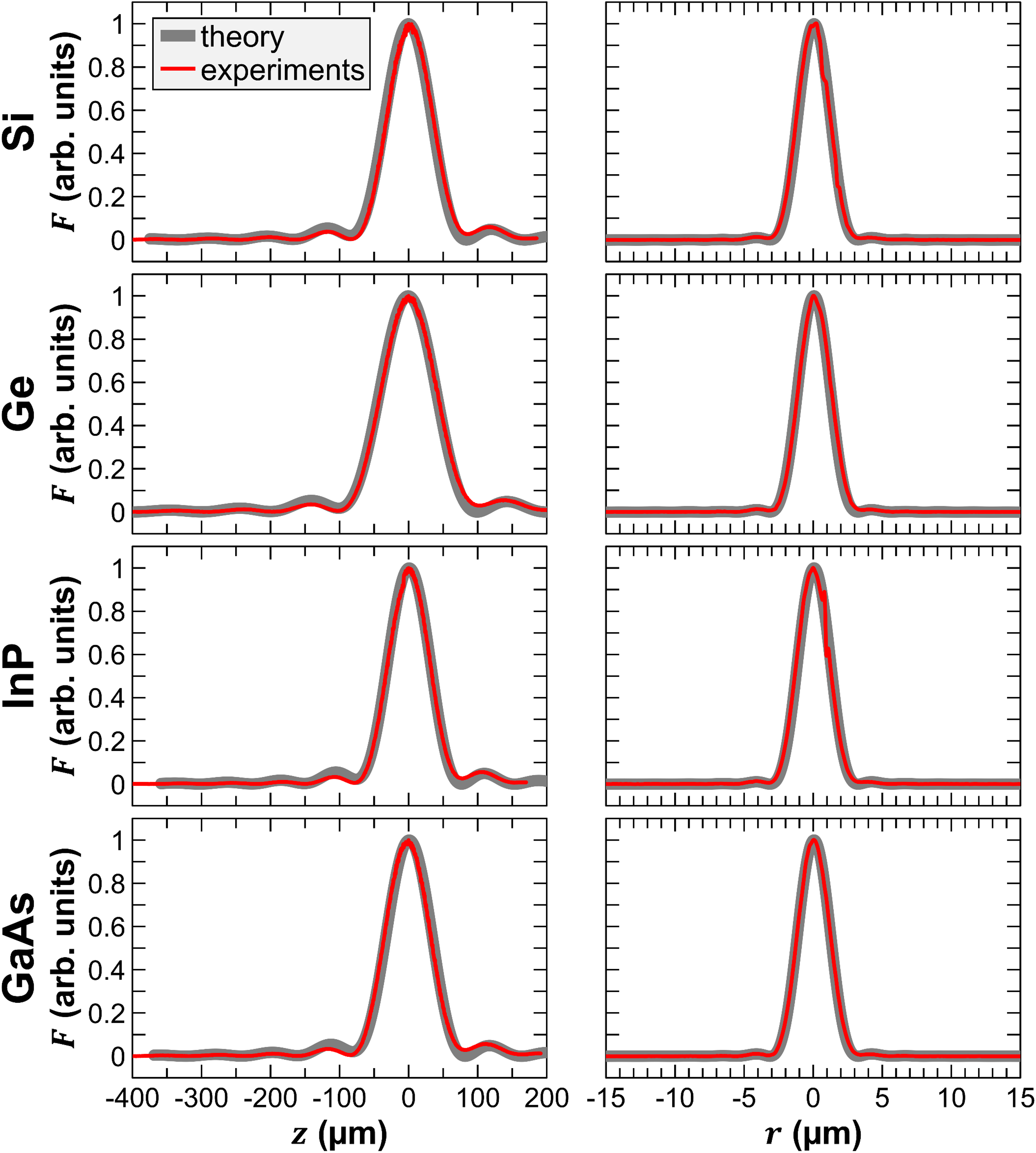}
\caption{\label{fig:FigSlinearregime} On-axis and radial fluence distributions in the linear regime for all semiconductors investigated. The gray curves are calculations with the vectorial model \textit{InFocus} \cite{Li2021,Li2021a} for a focusing depth of 500~$\mu$m inside the material, and the red curves are experimental measurements.}
\end{figure}

Generally speaking, spatial beam distortions can be caused by spherical aberration due to refractive index mismatch at the air--material interface. Let us examine how pronounced the spherical aberration is for each material. To do so, linear propagation calculations have been carried out with the vectorial model \textit{InFocus} \cite{Li2021,Li2021a}. The corresponding calculations of the spatial beam distribution along the optical axis $z$ and the radial direction $r$ are shown in Fig.~\ref{fig:FigSlinearregime} (gray curves). As expected, higher refractive indices $n_{0}$ lead to a more extended focal zone along $z$, and have no influence along $r$. For each material, the experimental measurements for the lowest $E_{\rm{in}}$ values (red curves) are in excellent agreement with the calculations. The corresponding theoretical and experimental beam waist ($w_{0}$) and Rayleigh length ($z_{R}$) values are recapitulated in Table~\ref{tab:w0zR}. For both parameters, the calculations underestimate the experimental values by $<3 \%$. This minor difference could originate from experimental uncertainties, as well as the assumption that the focusing lens does not cause additional aberrations. From the $F(r)$ profiles displayed in Fig.~\ref{fig:FigSlinearregime}, one can distinguish side peaks at around $\pm 4$~$\mu$m. The higher the amplitude of these peaks, the more pronounced the spherical aberration. Given that the amplitude of these peaks is $\approx 1\%$ of the main peak, we conclude that spherical aberration can be safely neglected, as additionally supported by the near symmetric $F(z)$ profiles.\\

\begin{table}[!ht]
  \centering
  \caption{Theoretical and experimental values for the beam radius at $1/e^{2}$ ($w_{0}^{\rm{th}}$ and $w_{0}^{\rm{exp}}$, respectively), and for the Rayleigh length ($z_{R}^{\rm{th}}$ and $z_{R}^{\rm{exp}}$, respectively).}
  \label{tab:w0zR}
  \scalebox{0.9}{\begin{tabular}{@{}ccccc@{}}
    \toprule
    Medium & $w_{0}^{\rm{th}}$ ($\mu$m) & $w_{0}^{\rm{exp}}$ ($\mu$m) & $z_{R}^{\rm{th}}$ ($\mu$m) & $z_{R}^{\rm{exp}}$ ($\mu$m)\\
    \midrule
    Si & 2.15 & 2.22 & 36.09 & 36.24\\
    Ge & 2.15 & 2.22 & 42.88 & 43.61\\
    InP & 2.15 & 2.20 & 32.79 & 33.09\\
    GaAs & 2.15 & 2.20 & 34.93 & 35.60\\
    \bottomrule
  \end{tabular}}
\end{table}

Increasing the numerical aperture $\text{NA}$ of the focusing optics leads to higher energy deposition in semiconductors \cite{Chanal2017}. However, even for aberration-corrected optical systems such as objective lenses, refraction at the air--solid interface with a high refractive index mismatch may cause severe spherical aberration. This aberration, which results in an asymmetric fluence distribution at the focus, is more pronounced for high refractive index mismatch, high $\text{NA}$, and a deep focus position inside the solid. To quantify the dependence of spherical aberration on NA, additional linear propagation calculations have been performed for different $\text{NA}$s with \textit{InFocus}, with similar conditions as in the experiments ($\lambda=1960$~nm, Gaussian beam with a diameter $\approx 60 \%$ larger than the entrance pupil of the focusing optics, linear polarization, sample thickness of 500~$\mu$m). The considered material is Ge, as it shows the highest refractive index ($n_{0}=4.106$) among the semiconductors investigated. This implies that, if the spherical aberration is not pronounced in Ge, the same is true for Si, InP, and GaAs. To evaluate how pronounced the spherical aberration is, we define the asymmetricity parameter $\sigma$ as
\begin{equation}
\label{eq:asymmetricity}
\sigma = \frac{1}{L} \left| \int_{0}^{z_{g}} F(z) dz - \int_{z_{g}}^{L} F(z) dz \right|,
\end{equation}
where $F$ is the on-axis fluence, $z_{g}$ is the on-axis position of the geometrical focus, and $L=500$~$\mu$m is the sample thickness. By construction, $\sigma \approx 0$ when spherical aberration is negligible.\\

\begin{figure}
\centering
\includegraphics[width=\linewidth]{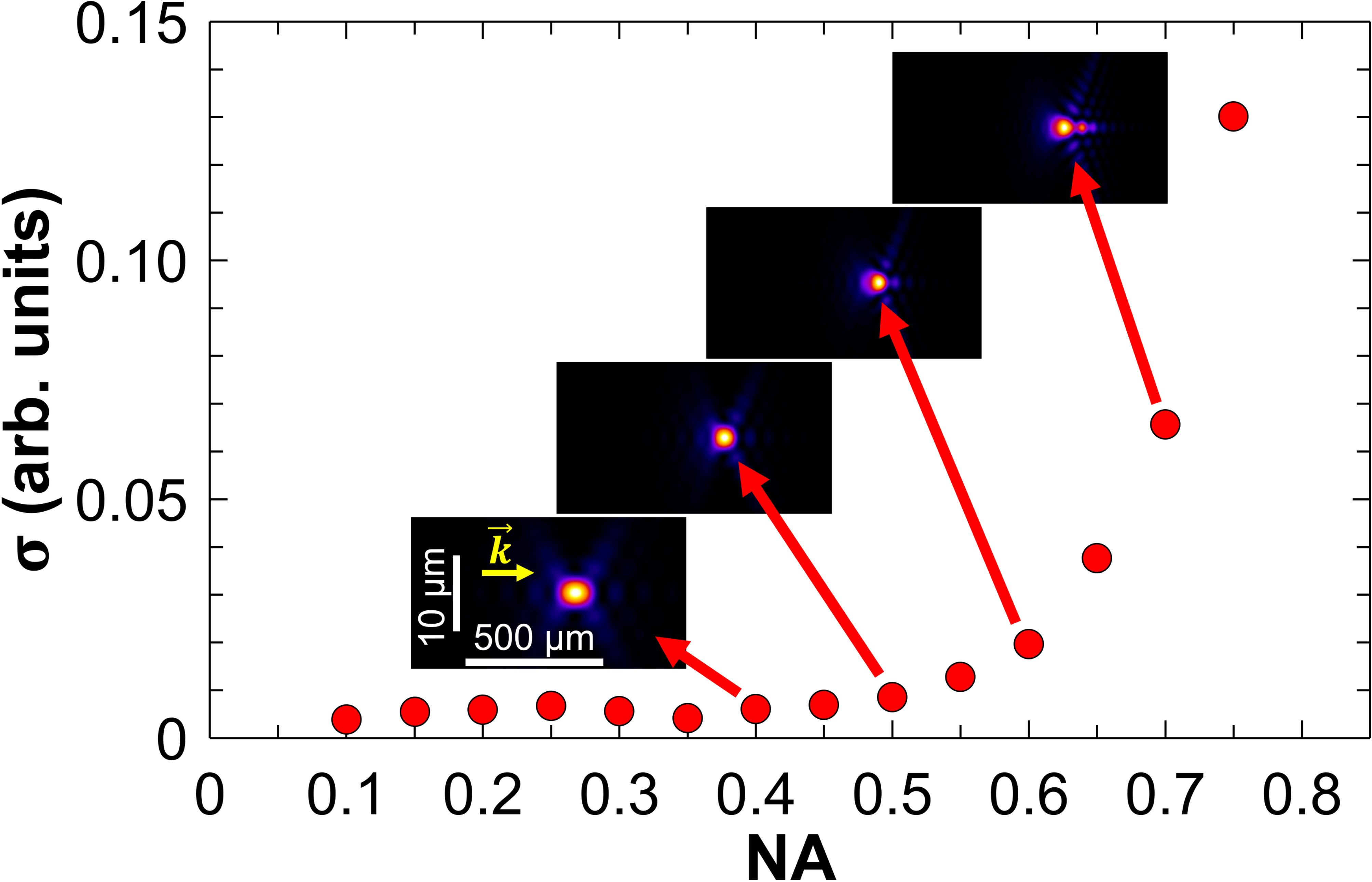}
\caption{\label{fig:FigSsphab} Evolution of the asymmetricity $\sigma$ in Ge as a function of the numerical aperture $\text{NA}$. The insets show the result of calculations with the vectorial model \textit{InFocus} \cite{Li2021,Li2021a} at different $\text{NA}$s. The vector $\vec{k}$ indicates the direction of laser propagation. The spatial scales apply to all images.}
\end{figure}

The evolution of $\sigma$ as a function of $\text{NA}$ is displayed in Fig.~\ref{fig:FigSsphab}. For $\text{NA} \le 0.45$, $\sigma \approx 0$, which demonstrates the absence of pronounced spherical aberration. In contrast, $\sigma$ increases for $\text{NA} \ge 0.50$, which indicates that spherical aberration can no longer be neglected. An important conclusion from the results in Fig.~\ref{fig:FigSsphab} is that the objective lens of $\text{NA} = 0.40$ used in all experiments is an excellent compromise to obtain tight focusing without pronounced spherical aberration.

\subsection{Maximum fluence in the linear regime}

In the linear propagation regime, the maximum fluence $F_{\rm{max}}$ reached in the bulk of the considered medium can be simply expressed as a function of the input pulse energy $E_{\rm{in}}$. Assuming a Gaussian fluence distribution
\begin{equation}
\label{eq:F(r)}
F(r)=F_{\rm{max}} e^{-\frac{2r^{2}}{w_{0}^2}},\\
\end{equation}
where $w_{0}$ is the beam radius at $1/e^{2}$, the input pulse energy $E_{\rm{in}}$ reads
\begin{equation}
\label{eq:Fmax_lin1}
\begin{split}
    T E_{\rm{in}} & =\int_{0}^{2\pi} \int_{0}^{+\infty} F(r) r dr d\theta\\
    & = 2\pi F_{\rm{max}} \int_{0}^{+\infty} r e^{-\frac{2r^{2}}{w_{0}^2}} dr\\
    & = 2\pi F_{\rm{max}} \frac{w_{0}^2}{4}.
\end{split}
\end{equation}
where $T=1-(1-n_{0})^{2}/(1+n_{0})^{2}$ is the Fresnel transmission coefficient at the air--medium interface at normal incidence. Thus, $F_{\rm{max}}$ reads
\begin{equation}
\label{eq:Fmax_lin2}
F_{\rm{max}}=\frac{2TE_{\rm{in}}}{\pi w_{0}^{2}}.
\end{equation}

As shown in the Primary Manuscript, Fig.~\ref{fig:Fig2}(d), calculations with Eq.~\eqref{eq:Fmax_lin2} are in excellent agreement with experimental measurements for low input pulse energy.

\subsection{Group velocity dispersion}
\label{ssec:GVD}

As shown in Table~\ref{tab:gvd}, semiconductors are highly dispersive optical media. Let us examine if group velocity dispersion in these media can cause temporal pulse broadening. For bandwidth-limited Gaussian pulses, the pulse duration $\tau_{\text{out}}$ exiting the sample reads
\begin{equation}
\label{eq:gvd}
    \tau_{\text{out}}=\tau_{\text{in}} \sqrt{1+\left( 4\ln{(2)}\text{GVD}\frac{d}{\tau_{\text{in}}^2} \right)^{2}}
\end{equation}
where $\tau_{\text{in}}$ is the pulse duration entering the sample, $\text{GVD}$ is the group velocity dispersion, and $d$ is the sample thickness.\\

\begin{table}[!ht]
  \centering
  \caption{Group velocity dispersion $\text{GVD}$ for all semiconductors investigated. The values are given for $\lambda \approx 1960$~nm.}
  \label{tab:gvd}
  \scalebox{1}{\begin{tabular}{@{}ccc@{}}
    \toprule
    Medium & $\rm{GVD}$ (fs$^{2}$/mm) & Reference\\
    \midrule
    Si & 830 & \cite{Salzberg1957}\\
    Ge & 3474 & \cite{Burnett2016}\\
    InP & 1101 & \cite{Pettit1965}\\
    GaAs & 1053 & \cite{Skauli2003}\\
    \bottomrule
  \end{tabular}}
\end{table}

The bandwidth limit calculated as the Fourier transform of the experimentally measured spectrum is $\tau_{\text{in}}=257$~fs.
The difference between this theoretical limit and the minimum pulse duration of 275~fs determined with autocorrelation originates from higher-order dispersion. Using the $\text{GVD}$ values in Table~\ref{tab:gvd} and the sample thickness of $d=500$~$\mu$m, the temporal pulse broadening $\Delta \tau = \tau_{\text{out}}-\tau_{\text{in}}$ according to Eq.~\eqref{eq:gvd} is $<1$~fs for all media. We thus conclude that group velocity dispersion can be safely neglected in all our experiments. This holds all the more for experiments at longer pulse durations. An important conclusion is that the differences observed in Si for up- and down-chirped 3-ps pulses [Primary Manuscript, Fig.~\ref{fig:Fig4}(a)--(c)] cannot originate from group velocity dispersion.

\section{Filamentation regime}
\label{sec:filamentationregime}

\subsection{Propagation morphology}
\label{ssec:morpho}

\begin{figure}
\centering
\includegraphics[width=\linewidth]{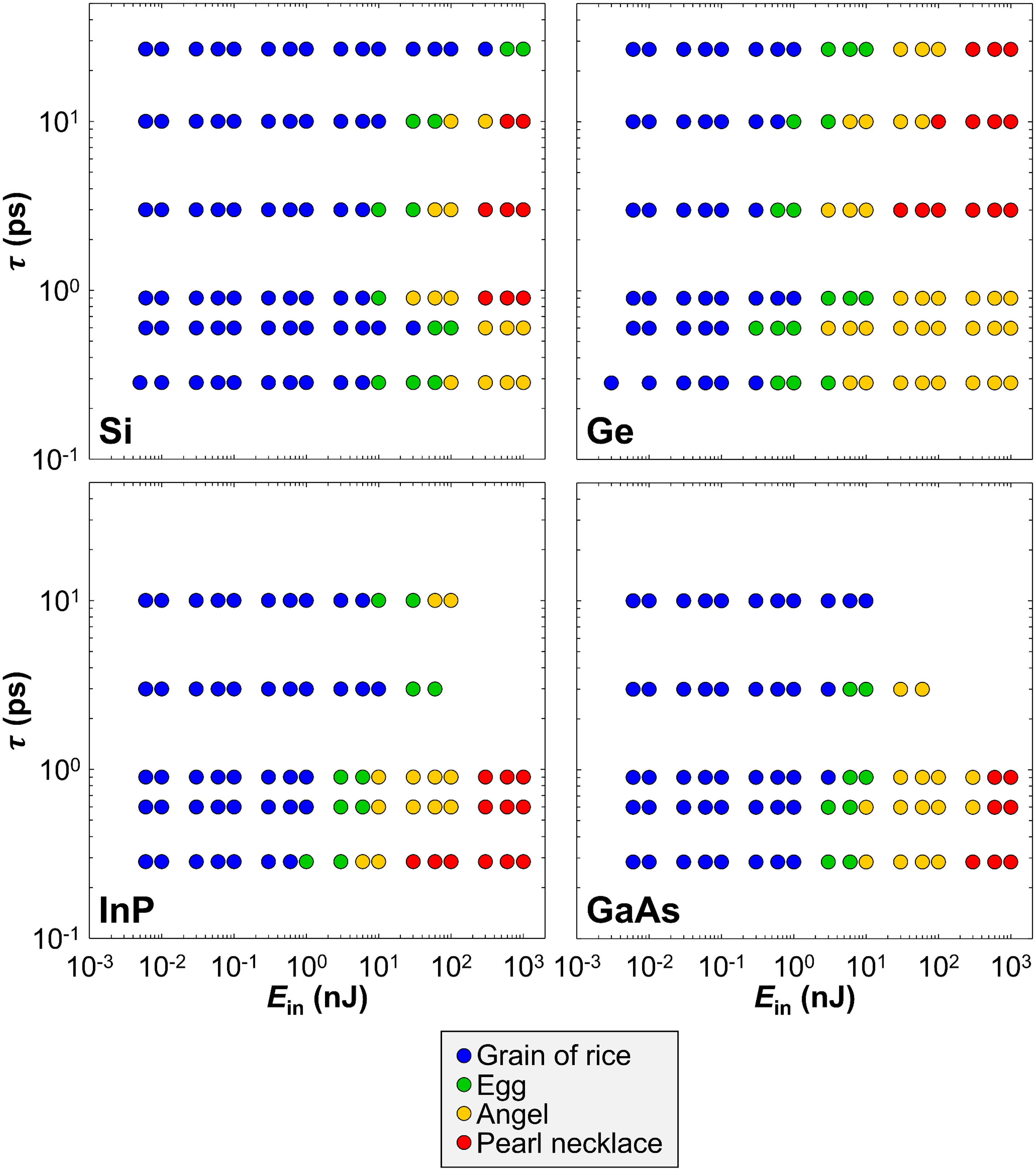}
\caption{\label{fig:FigSmorphodomains} Morphology of the fluence distributions in all semiconductors investigated for different input pulse energy $E_{\rm{in}}$ and pulse duration $\tau$.}
\end{figure}

As exemplified in the Primary Manuscript, Fig.~\ref{fig:Fig2}(c), the morphology of the fluence distribution depends on the input pulse energy $E_{\rm{in}}$. The morphologies obtained for different $E_{\rm{in}}$ and $\tau$ conditions have been examined for all semiconductors investigated (Fig.~\ref{fig:FigSmorphodomains}). A common feature for all materials is that, for increased $E_{\rm{in}}$, the propagation morphology changes consecutively from \textit{grain of rice} to \textit{egg} to \textit{angel} to \textit{pearl necklace}. Nevertheless, the transition $E_{\rm{in}}$ value between two morphologies strongly depends on the considered medium and pulse duration. This is in excellent agreement with the results shown in the Primary Manuscript, Fig.~\ref{fig:Fig3}, where nonlinear refraction and absorption depend on $\tau$.\\

\subsection{Energy profiles}
\label{ssec:Ez}

The evolution of the energy $E$ as a function of the on-axis distance $z$ [i.e., the $E(z)$ profile] is obtained by integrating the 3D fluence distributions $F(x,y,z)$ over $x$ and $y$
\begin{equation}
\label{eq:Ez}
E(z)=\iint_{\mathbb{R}^{2}} F(x,y,z) dx dy.\\
\end{equation}

\begin{figure}
\centering
\includegraphics[width=\linewidth]{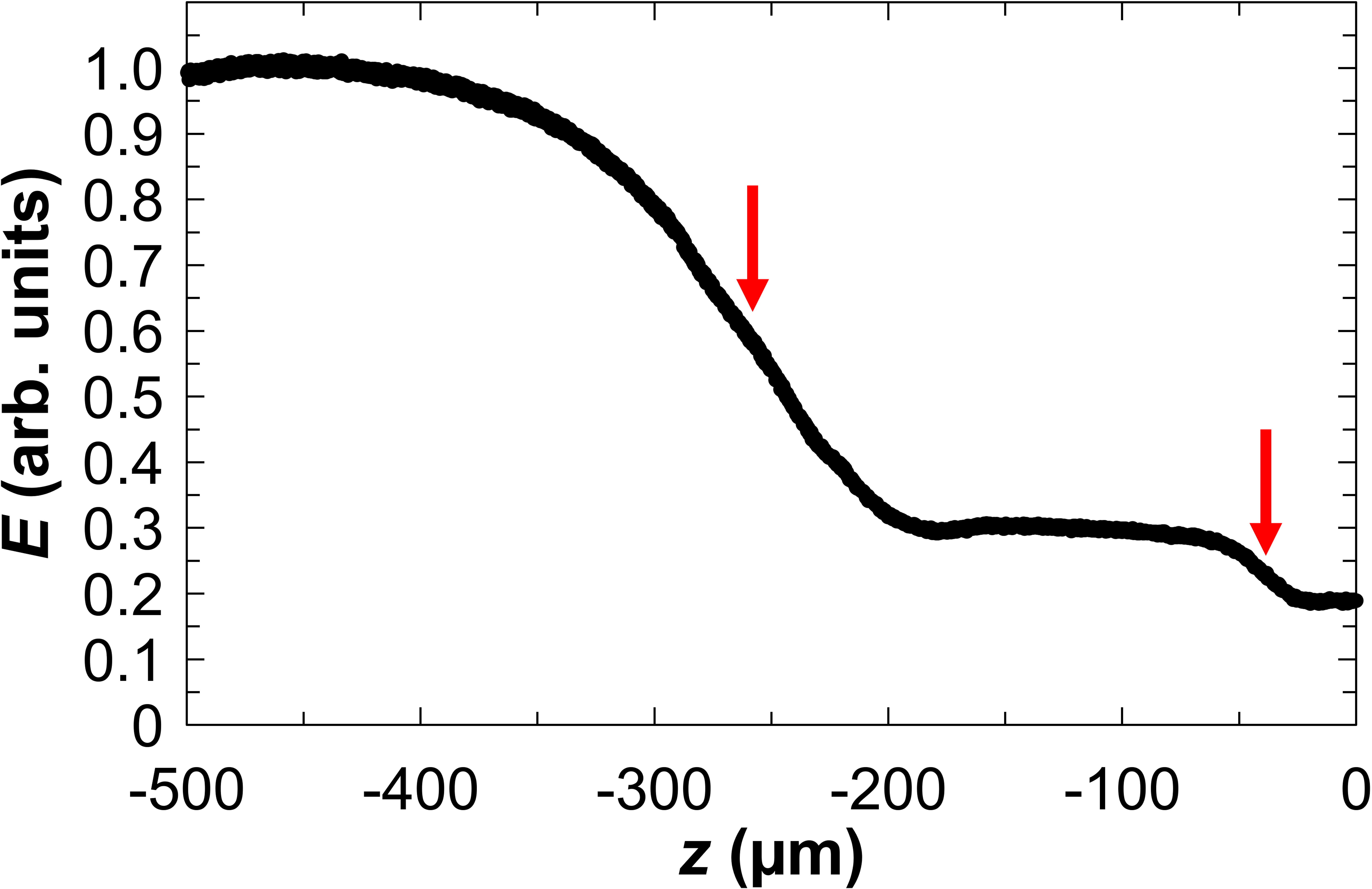}
\caption{\label{fig:FigSEzlimit} Evolution of the normalized energy $E$ in Si for $E_{\rm{in}}=1$~$\mu$J and $\tau=3$~ps as a function of the on-axis distance $z$. The red arrows indicate the different inflection points for the same $E(z)$ profile.}
\end{figure}

As shown in the Primary Manuscript, Fig.~\ref{fig:Fig3}(e) and (f), the $E(z)$ profiles are well-described by sigmoid functions defined as
\begin{equation}
\label{eq:sigmoid}
  E(z)=A-\frac{B}{1+\exp{(-\frac{z-D}{C})}}.
\end{equation}
The fitting parameters $A$ and $B$ define the energy before and after the interaction as $E(-\infty)=T E_{\rm{in}}=A$ and $E(+\infty)=A-B$, respectively. The fraction of absorbed energy is defined as $f_{E}=1-E(+\infty)/E(-\infty)=B/A$. The steepness of the sigmoid is inversely proportional to the parameter $C$, which can be defined as the characteristic absorption length ($C=L_{\rm{abs}})$. Finally the parameter $D$ defines the position of the inflection point. A necessary fitting condition is that the $E(z)$ profile must exhibit a single inflection point. A counter-example for high $E_{\rm{in}}$ is shown in Fig.~\ref{fig:FigSEzlimit}, where two inflection points exist, invalidating the sigmoid fit defined in Eq.~\eqref{eq:sigmoid}. This situation of multiple inflection points on the $E(z)$ profile is generally obtained at high energy, where the filament exhibits an \textit{angel} or \textit{pearl necklace} morphology (see Section \ref{ssec:morpho}).\\

As shown in the Primary Manuscript, Fig.~\ref{fig:Fig3}(g) and (h), $f_{E}$ and $L_{\rm{abs}}$ both scale logarithmically with $E_{\rm{in}}$, and also with $1/\tau$. This has been examined by fitting the experimental data as
\begin{equation}
\label{eq:fitfeLabs}
  \begin{cases}
    f_{E}=a_{f_{E}}\ln{(\tau)}+b_{f_{E}}\ln({E_{\rm{in}}})+c_{f_{E}},\\
    L_{\rm{abs}}=a_{L_{\rm{abs}}}\ln{(\tau)}+b_{L_{\rm{abs}}}\ln({E_{\rm{in}}})+c_{L_{\rm{abs}}}.
  \end{cases}
\end{equation}
The corresponding fitting coefficients are given in Table~\ref{tab:logfit}. The striking feature highlighted by the nearly parallel planes in the Primary Manuscript, Fig.~\ref{fig:Fig3}(g) is that the coefficients $a_{f_{E}}$ and $b_{f_{E}}$ are very similar for all materials. The material-dependent coefficient $c_{f_{E}}$ represents the threshold for energy absorption. In contrast, all fitting coefficients for $L_{\rm{abs}}$ strongly differ from one material to another.

\begin{table}[!ht]
  \centering
  \caption{Fitting coefficients $a_{f_{E}}$, $b_{f_{E}}$, $c_{f_{E}}$, $a_{L_{\rm{abs}}}$, $b_{L_{\rm{abs}}}$, and $c_{L_{\rm{abs}}}$ in Eq.~\eqref{eq:fitfeLabs} for all tested semiconductors.}
  \label{tab:logfit}
  \scalebox{0.76}{\begin{tabular}{@{}ccccc@{}}
    \toprule
    Medium & Si & Ge & InP & GaAs\\
    \midrule
    $a_{f_{E}}$ & $-0.11 \pm 0.01$ & $-0.15 \pm 0.01$ & $-0.14 \pm 0.01$ & $-0.12 \pm 0.02$\\
    $b_{f_{E}}$ & $0.16 \pm 0.01$ & $0.16 \pm 0.01$ & $0.15 \pm 0.01$ & $0.17 \pm 0.01$\\
    $c_{f_{E}}$ & $-0.17 \pm 0.03$ & $0.57 \pm 0.02$ & $-0.04 \pm 0.03$ & $-0.05 \pm 0.04$\\
    $a_{L_{\rm{abs}}}$ & $-3.13 \pm 0.88$ & $-8.09 \pm 1.70$ & $-4.42 \pm 1.13$ & $-6.72 \pm 1.87$\\
    $b_{L_{\rm{abs}}}$ & $3.30 \pm 1.18$ & $8.56 \pm 1.32$ & $4.65 \pm 1.06$ & $7.82 \pm 1.27$\\
    $c_{L_{\rm{abs}}}$ & $1.73 \pm 4.65$ & $35.23 \pm 2.12$ & $-0.94 \pm 3.45$ & $-7.67 \pm 4.55$\\
    \bottomrule
  \end{tabular}}
\end{table}

\subsection{Critical power for nonlinearities}
\label{ssec:Pcr}

\textbf{Experimental determination}\\

\begin{figure}[!ht]
\centering
\includegraphics[width=\linewidth]{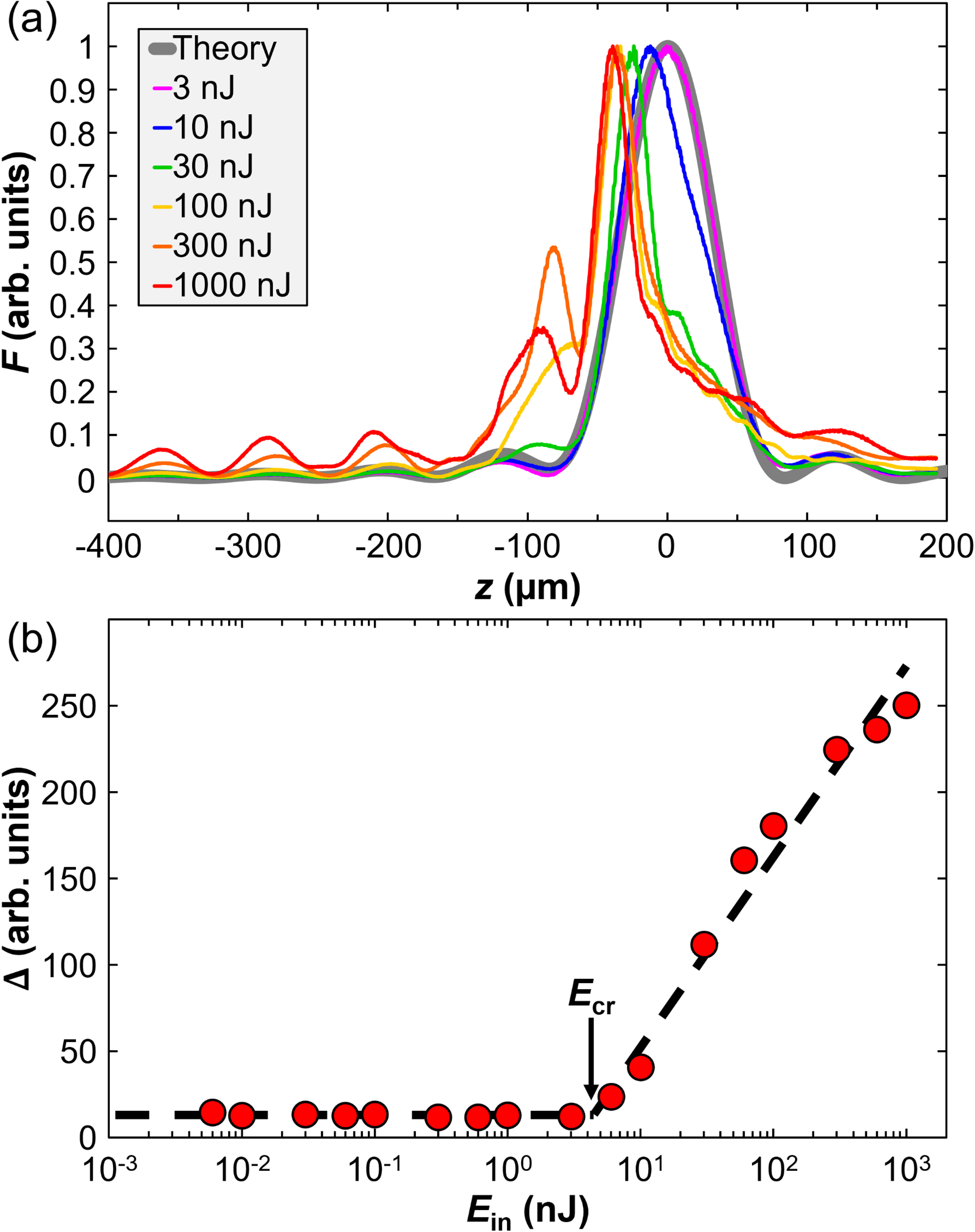}
\caption{\label{fig:FigSdeviationlinear} Determination of the critical pulse energy $E_{\rm{cr}}$ which delimits the linear and the nonlinear propagation regimes. (a) On-axis normalized fluence distributions in Si for different input pulse energies $E_{\rm{in}}$. The gray curve has been calculated in the linear propagation regime with the vectorial model \textit{InFocus} \cite{Li2021,Li2021a}, and the other curves have been obtained experimentally for $\tau=900$~fs. (b) Evolution of $\Delta$ according to Eq.~\eqref{eq:deltalinearregime} between theory and experiments as a function of the input pulse energy $E_{\rm{in}}$. The dashed lines are guides for the eye.}
\end{figure}

The on-axis distributions calculated in Fig.~\ref{fig:FigSlinearregime} serve as a benchmark for determining the input pulse energy $E_{\rm{cr}}$, which delimits the linear and the nonlinear propagation regimes. As exemplified in Fig.~\ref{fig:FigSdeviationlinear}(a), the on-axis fluence profiles are determined for various $E_{\rm{in}}$ values, and directly compared to calculations performed with \textit{InFocus} (see Section~\ref{sec:linpropregime}). For pulse energies $E_{\rm{in}} \le 3$~nJ, the measurements are in excellent agreement with calculations. However, for higher $E_{\rm{in}}$ values, the experimental profiles deviate from the linear regime due to Kerr and plasma effects. We introduce the difference $\Delta$ between the experimental and theoretical profiles as
\begin{equation}
\label{eq:deltalinearregime}
\Delta=\int_{-\infty}^{+\infty} \left| F^{\rm{exp}}(z)-F^{\rm{th}}(z) \right| dz,
\end{equation}
where $F^{\rm{exp}}$ and $F^{\rm{th}}$ are the experimental and theoretical on-axis fluence profiles, respectively. The model \textit{InFocus} is used to calculate $F^{\rm{th}}$ with $250$-nm steps along the $z$-axis.\\

An example of the evolution of $\Delta$ as a function of $E_{\rm{in}}$ is displayed in Fig.~\ref{fig:FigSdeviationlinear}(b). For $E_{\rm{in}} \le 3$~nJ, $\Delta$ is nearly constant, again highlighting that the propagation regime is linear. In contrast, $\Delta$ increases with $E_{\rm{in}}>3$~nJ, which indicates that the propagation regime is nonlinear. The critical pulse energy $E_{\rm{cr}}$, which delimits the two propagation regimes is thus determined as the average between the highest input pulse energy for which the propagation is linear [$E_{\rm{in}}=3$~nJ in Fig.~\ref{fig:FigSdeviationlinear}(b)], and the lowest input pulse energy for which the propagation is nonlinear [$E_{\rm{in}}=6$~nJ in Fig.~\ref{fig:FigSdeviationlinear}(b)]. The effective critical power shown in the Primary Manuscript, Fig.~\ref{fig:Fig3}(b) is evaluated for different pulse durations $\tau$ as
\begin{equation}
\label{eq:Pcr}
  P_{\rm{cr}}^{\rm{eff}}=0.88 \frac{T E_{\rm{cr}}}{\tau},
\end{equation}
where $T$ is the Fresnel transmission coefficient.\\

\noindent \textbf{Temporal scaling laws}\\

In the Primary Manuscript, we have shown that the measured critical power $P_{\rm{cr}}^{\rm{eff}}$ decreases for increasing pulse duration $\tau$ [see Fig.~\ref{fig:Fig3}(b)]. To give a physical interpretation of this temporal dependence, let us consider the nonlinear polarization $P_{\rm{nl}}$. Considering only the Kerr effect, and neglecting third harmonic generation, $P_{\rm{nl}}$ can be written as
\begin{align}
\label{eq:Pnl}
  P_{\rm{nl}}(t) = 2 n_{0} \varepsilon_{0} n_{2}^{\rm{eff}}(t) I(t) E(t),
\end{align}
where $\varepsilon_{0}$ is the vacuum permittivity, $n_{0}$ is the linear refractive index, $I(t)$ is the intensity, $E(t)$ is the magnitude of the electric field, and $n_{2}^{\rm{eff}}(t)$ is the time-dependent effective nonlinear index. The temporal dependence of $n_{2}^{\rm{eff}}(t)$ implies a delayed medium response, which can be written in the most generic form as
\begin{align}
\label{eq:n2eff}
  n_{2}^{\rm{eff}}(t)
  = n_{2} \frac{\int_{-\infty}^t H(t-t') I(t') dt'}{I(t)}
  = n_{2} \frac{H \otimes I}{I(t)},
\end{align}
where $H(t)$ is the medium response function which must satisfy the causality condition $H(t<0)=0$, and be normalized to unity so that $\int_{0}^\infty H(t) dt = 1$. In Eq.~\eqref{eq:n2eff}, $n_{2}$ is a nonlinear index corresponding to an infinitely long laser pulse compared to the characteristic medium response time $\tau_{r}$. Indeed, for $\tau \gg \tau_{r}$, $H(t)$ can be considered as a Dirac delta function, resulting in $H \otimes I = I$, which leads to $n_{2}^{\rm{eff}}=n_{2}$.\\

Let us first consider a Gaussian medium response
\begin{align}
\label{eq:Hgauss}
H(t) = \frac{2\eta}{\sqrt{\pi}\tau_{r}} \exp\left(-\eta^{2}\frac{t^{2}}{\tau_{r}^{2}}\right),
\end{align}
where $\tau_{r}$ is the characteristic response time of the medium, and $\eta$ is a shape factor. Below, we set $\eta=2\sqrt{\ln{2}}$, so that $\tau_{r}$ is defined at FWHM. Furthermore, let us assume a Gaussian intensity profile $I(t)$
\begin{align}
\label{eq:I}
  I(t) = I_{0} \exp{\left( - \eta^{2}\frac{t^{2}}{\tau^{2}} \right)},
\end{align}
where $I_{0}$ is the peak intensity, and $\tau$ is the duration at FWHM. Using Eqs.~\eqref{eq:Hgauss} and \eqref{eq:I}, the convolution in Eq.~\eqref{eq:n2eff} can be expressed analytically as
\begin{align}
\label{eq:convolutiongaussian}
H \otimes I = \frac{I_{0}}{\sqrt{1+\frac{\tau_{r}^{2}}{\tau^{2}}}}\exp{\left( -\eta^{2}\frac{t^{2}}{\tau_{r}^{2} + \tau^{2}}\right)} \notag \\
\mathrm{erfc}\left(-\eta\frac{\tau_{r}}{\tau}\frac{t}{\sqrt{\tau_{r}^{2} + \tau^{2}}}\right),
\end{align}
where $\mathrm{erfc}=1-\mathrm{erf}$ is the complementary error function. From Eqs.~\eqref{eq:Pnl}, \eqref{eq:n2eff} and \eqref{eq:I}, one can qualitatively conclude that the propagation is mainly influenced by the effective nonlinear refractive index $n_{2}^{\rm{eff}}$ at $t=0$ for which the intensity reaches its maximum value $I_{0}$---and thus, the nonlinear polarization $P_\text{nl}$ is the strongest. Combining Eqs.~\eqref{eq:n2eff} and \eqref{eq:convolutiongaussian} at $t=0$, the effective nonlinear index reads
\begin{align}
    n_{2}^{\text{eff}}(t=0) = \frac{n_{2}}{\sqrt{1+\frac{\tau_{r}^{2}}{\tau^{2}}}},
\end{align}
which leads to the following simple expression for the effective critical power $P_\text{cr}^\text{eff}$ as the function of $\tau$
\begin{align}
\label{eq:Pcr_vs_tau0_gaussian}
    P_{\text{cr}}^{\text{eff}}(\tau) = P_{\text{cr}} \sqrt{1+\frac{\tau_{r}^{2}}{\tau^{2}}},
\end{align}
where $P_{\rm{cr}} = \alpha \lambda^{2}/(4\pi n_{0} n_{2})$ is the critical power obtained for long pulses for which the effect of the delayed medium response is negligible, and $\alpha=1.8962$.\\

\begin{figure}[!ht]
  \includegraphics[width=\linewidth]{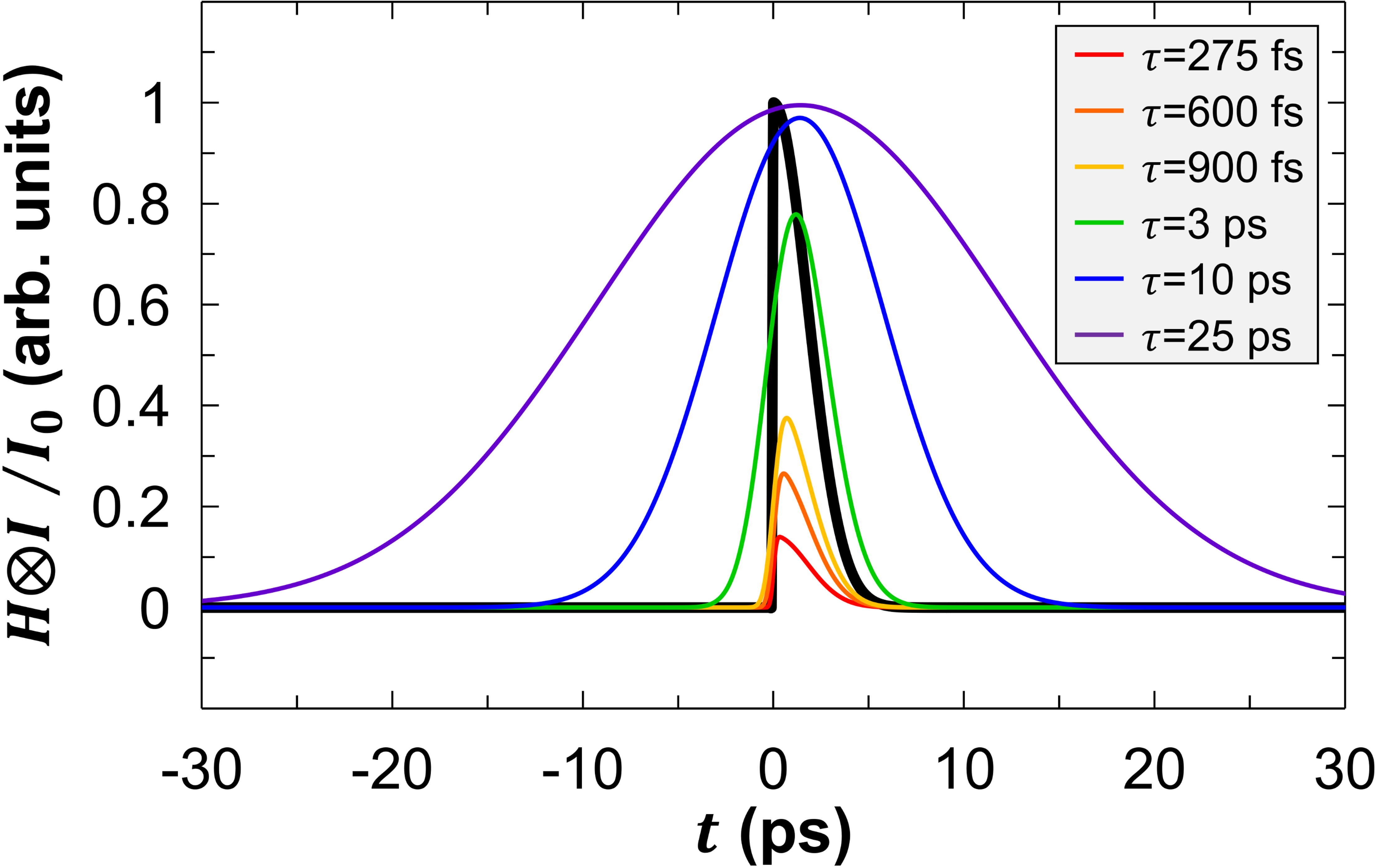}
  \caption{\label{fig:FigSconvolution}
    Time dependence of the normalized convolution integral $H \otimes I / I_{0}$ according to Eq.~\eqref{eq:convolutiongaussian} for the experimental pulse durations $\tau$. The black curve is the normalized Gaussian response function $H(t)$ [Eq.~\eqref{eq:Hgauss}], with a response time $\tau_{r}=4.2$~ps.
  }
\end{figure}

The temporal evolution of the normalized convolution $H \otimes I/I_{0}$ according to Eq.~\eqref{eq:convolutiongaussian} is shown in Fig.~\ref{fig:FigSconvolution} for the experimental pulse durations $\tau$ investigated. The medium response is assumed to be Gaussian, with a characteristic time $\tau_{r}=4.2$~ps corresponding to the average value for all semiconductors investigated (see below). The value of $H \otimes I/I_{0}$ at $t=0$---and thus, the effective nonlinear refractive index $n_{2}^{\rm{eff}}$---increases with the pulse duration. As $P_{\rm{cr}}^{\rm{eff}} \propto 1/n_{2}^{\rm{eff}}$, one can conclude that the effective critical power decreases with the pulse duration, in good agreement with the experimental results in the Primary Manuscript, Fig.~\ref{fig:Fig3}(b). While quantitative differences may arise when selecting different shapes for the response function $H(t)$, it does not qualitatively impact on the trend that $P_{\rm{cr}}^{\rm{eff}}$ decreases with $\tau$.\\

For the same Gaussian intensity profile $I(t)$ defined in Eq.~\eqref{eq:I}, one could also consider an exponentially decaying response function
\begin{align}
\label{eq:Hexp}
  H(t) = \frac{\eta}{\tau_{r}} \exp{\left( -\eta \frac{t}{\tau_{r}} \right)},
\end{align}
In this case the convolution in Eq.~\eqref{eq:n2eff} becomes
\begin{align}
\label{eq:convolution}
  H \otimes I = I_0 \frac{\sqrt{\pi}}{2} \frac{\tau}{\tau_{r}} \exp{\left({\frac{1}{4}\frac{\tau^{2}}{\tau_{r}^{2}} -\eta\frac{t}{\tau_{r}}}\right)} \notag \\
  \mathrm{erfc}\left( \frac{1}{2}\frac{\tau}{\tau_{r}}-\frac{\eta^{2}}{2}\frac{t}{\tau}\right),
\end{align}
and the corresponding effective nonlinear index $n_{2}^{\text{eff}}$ at $t=0$ becomes
\begin{align}
n_{2}^\text{eff}(t=0)= n_{2} \frac{\sqrt{\pi}}{2}\frac{\tau}{\tau_{r}} \exp{ \left( \frac{1}{4}\frac{\tau^{2}}{\tau_{r}^{2}} \right) }
\mathrm{erfc}\left( \frac{1}{2}\frac{\tau}{\tau_{r}}\right).
\end{align}
Therefore, the effective critical power $P_\text{cr}^\text{eff}$ takes the following form
\begin{align}
\label{eq:Pcr_vs_tau0_expdecay}
P_{\rm{cr}}^{\rm{eff}}(\tau) = P_{\rm{cr}} \frac{2}{\sqrt{\pi}} \frac{\tau_{r}}{\tau} \frac{\exp{\left( -\frac{1}{4}\frac{\tau^{2}}{\tau_{r}^{2}} \right)}}{\mathrm{erfc} \left( \frac{1}{2}\frac{\tau}{\tau_{r}} \right)}.
\end{align}

To estimate the response time $\tau_{r}$ for different semiconductors, nonlinear curve fitting of the experimental data presented in the Primary Manuscript, Fig.~\ref{fig:Fig3}(b) is applied for a Gaussian response using Eq.~\eqref{eq:Pcr_vs_tau0_gaussian}. The chosen values for $P_{\rm{cr}}$ correspond to the effective critical powers determined with the longest pulses with which measurements could be performed. The obtained fits are superimposed with the experimental $P_{\rm{cr}}^{\rm{eff}}$ values in the Primary Manuscript, Fig.~\ref{fig:Fig3}(b). For all semiconductors, the trends given by Eq.~\eqref{eq:Pcr_vs_tau0_gaussian} are well reproduced, allowing us to conclude that the major physical mechanisms responsible for the decrease of $P_{\rm{cr}}^{\rm{eff}}$ with increasing pulse duration $\tau$ are correctly described. Similar results (not shown here) have been obtained using Eq.~\eqref{eq:Pcr_vs_tau0_expdecay} assuming an exponentially decaying medium response. As shown in Table~\ref{tab:taur}, for both considered response shapes, the response times $\tau_{r}$ for all materials are on the same order of magnitude.

\begin{table}[!ht]
  \centering
  \caption{
    Medium response time $\tau_{r}$ (in picoseconds) obtained by nonlinear curve fitting of the experimental $P_{\text{cr}}^{\text{eff}}$ data for Gaussian [Eq.~\eqref{eq:Hgauss}], and exponentially decaying [Eq.~\eqref{eq:Hexp}] response function $H(t)$.}
  \label{tab:taur}
  \begin{tabular}{@{}ccccc@{}}
    \toprule
     $H(t)$                   & Si   & Ge    & InP  & GaAs \\
    \midrule
     Gaussian                 & 2.94 & 11.41 & 1.72 & 0.67 \\
     exponential decay        & 2.22 & 9.19  & 1.28 & 0.47 \\
    %
    \bottomrule
  \end{tabular}
\end{table}

\subsection{Multi-photon absorption coefficient}
\label{ssec:MPA}

To determine the $N$-photon absorption coefficients $\beta_{N}$, we apply our recently developed approach relying on a modified Marburger formula, where power losses are accounted for \cite{Chambonneau2021}. The nonlinear focal shift $\Delta z$ reads
\begin{equation}
\label{eq:Marburger}
  \Delta z= - \left( d-\frac{1}{1/d+1/z_{\text{nl}}} \right),
\end{equation}
where $d=500$~$\mu$m is the sample thickness, and $z_{\text{nl}}$ is given by Marburger formula \cite{Marburger1975}
\begin{equation}
\label{eq:znl}
  z_{\text{nl}} = \frac{0.367k_{0}a_{0}^{2}}{\sqrt{ \left( \sqrt{P/P_{\text{cr}}} - 0.852 \right)^{2} -0.0219 }},
\end{equation}
where $k_{0}=n_{0}\omega_{0}/c$ is the wave number with $n_{0}$, $\omega_{0}$ and $c$ corresponding to the linear refractive index, the pulse central angular frequency, and the speed of light in vacuum, respectively, $a_{0}=d \tan{(\arcsin{(\text{NA}/n_{0})})}$ is the beam radius at the entrance of the sample, $\text{NA}=0.40$ is the numerical aperture of the focusing lens, and $P$ is the peak power. Here, $P_{\text{cr}}$ is taken as the effective critical power determined with the method described in Section~\ref{ssec:Pcr}.\\

To evaluate the peak power $P$ at the exit surface after propagation losses, we use the time-independent nonlinear Schrödinger equation
\begin{equation}
\label{eq:tindepNLSE}
  \frac{\partial E}{\partial z} = \frac{i}{2k_{0}}\Delta_{\perp}E - \frac{\beta_{N}}{4}I^{N-1}E,
\end{equation}
where $E$ is the magnitude of the electric field, and $I$ is the intensity. The first and second terms on the right-hand side of Eq.~\eqref{eq:tindepNLSE} correspond to diffraction and $N$-photon absorption, respectively. Given the tight focusing conditions that we use, the dominant propagation mechanism is diffraction, and multi-photon absorption mainly leads to a decrease of the peak intensity without altering the propagation. The assumption that diffraction and multi-photon absorption act independently should hold true as long as the peak power $P$ does not exceed the critical power $P_{\rm{cr}}$ by several orders of magnitude. In this regime, one can safely assume that the pulse duration $\tau$ is constant. Under these assumptions, one can define the electric fields $E_{d}$ and $E_{a}$ (the subscripts $d$ and $a$ standing for diffraction and absorption, respectively), which satisfy
\begin{equation}
\label{eq:Ed}
\frac{\partial E_{d}}{\partial z}=\frac{i}{2k_{0}}\Delta_{\perp}E_{d},\\
\end{equation}
and
\begin{equation}
\label{eq:Ea}
\frac{\partial E_{a}}{\partial z}=- \frac{\beta_{N}}{4}I^{N-1}E_{a}.
\end{equation}

Multiplying Eqs.~\eqref{eq:Ed} and \eqref{eq:Ea} by the complex conjugates $E_{d}^{*}$ and $E_{a}^{*}$, respectively, the solutions $I_{d}$ and $I_{a}$ for the differential equations read
\begin{equation}
\label{eq:Id}
I_{d}(x,y,z)=\frac{a_{0}^{2}}{a(z)^{2}}\exp{\left( {-\frac{x^{2}+y^{2}}{a(z)^{2}}} \right)},
\end{equation}
and
\begin{equation}
\label{eq:Ia}
I_{a}(z)=\frac{I_{0}}{\left[ 1+\frac{N-1}{2}\beta_{N}I_{0}^{N-1}z \right]^{\frac{1}{N-1}}},
\end{equation}
where $a(z)$ is the beam radius at the on-axis position $z$.\\

Finally, the peak power $P(z)$ can be expressed as a function of the intensity $I=I_{d}I_{a}$ as
\begin{equation}
\label{eq:P(z)}
\begin{split}
    P(z) & = \iint_{\mathbb{R}^{2}} I(x,y,z) dx dy\\
    & = \frac{a_{0}^{2}}{a(z)^{2}}I_{a}(z) \iint_{\mathbb{R}^{2}} \exp{\left( {-\frac{x^{2}+y^{2}}{a(z)^{2}}} \right)} dx dy\\
    & = \pi a_{0}^{2} I_{a}(z)\\
    & = \frac{P_{0}}{\left[ 1+\frac{N-1}{2}\beta_{N}\left( \frac{P_{0}}{\pi a_{0}^{2}} \right)^{N-1}z \right]^{\frac{1}{N-1}}},
\end{split}
\end{equation}
where $P_{0}=I_{0}\pi a_{0}^{2}$ is the power right after the entrance surface of the material considered. To extract $\beta_{N}$, nonlinear curve fitting of experimental $\Delta z$ data is applied using Eqs.~\eqref{eq:Marburger}, \eqref{eq:znl} where $P_{\rm{cr}}^{\rm{eff}}$ is determined following the method detailed in Section~\ref{ssec:Pcr}, and \eqref{eq:P(z)}.\\

\begin{figure}[!ht]
\centering
\includegraphics[width=\linewidth]{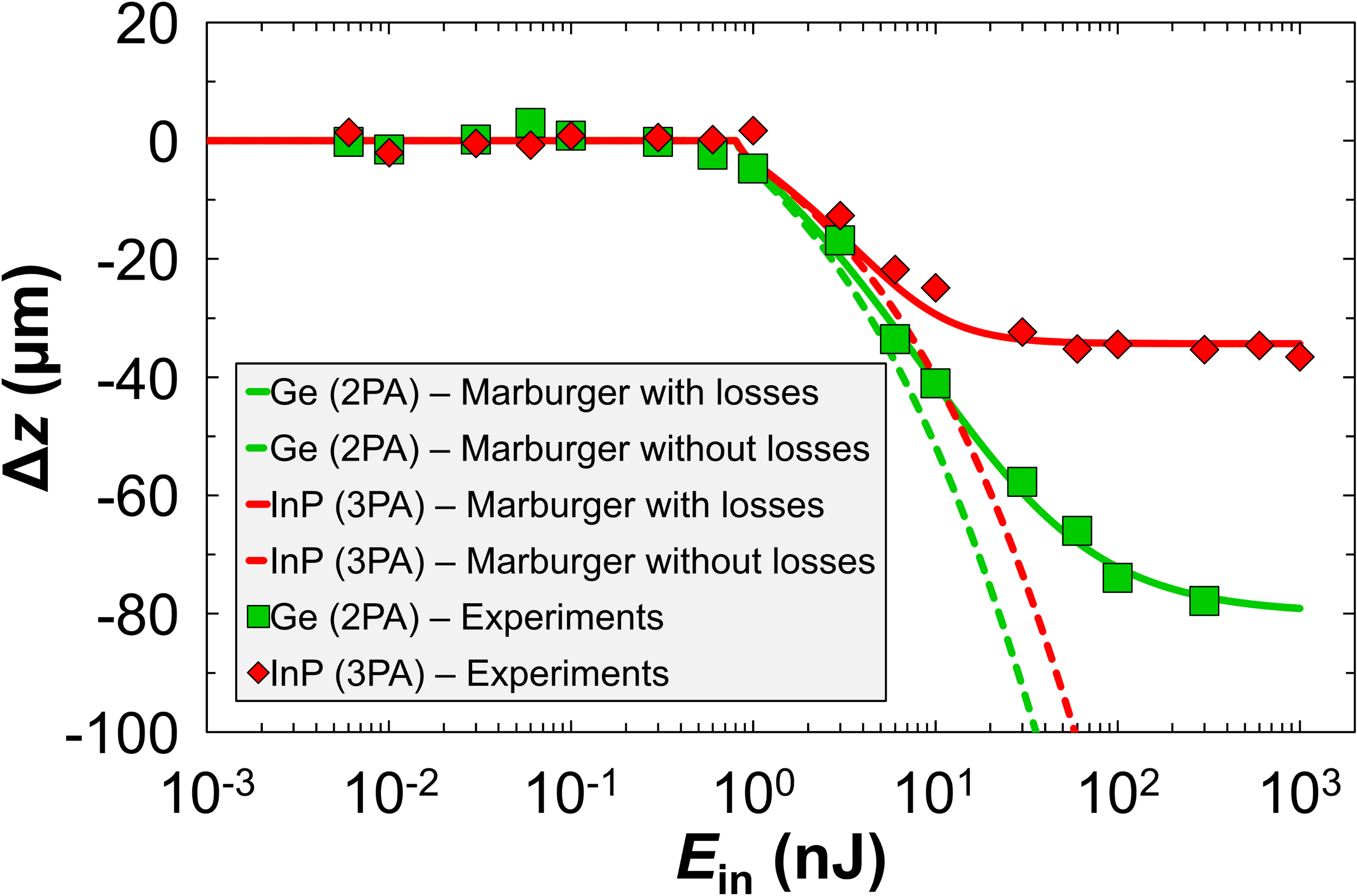}
\caption{\label{fig:FigSMarburger} Evolution of the nonlinear focal shift $\Delta z$ as a function of the input pulse energy $E_{\rm{in}}$ for Ge (green, 2PA, $\tau=900$~fs) and InP (red, 3PA, $\tau=10$~ps). The solid and dashed curves are calculations according to Marburger formulas accounting for and neglecting nonlinear power losses, respectively.}
\end{figure}

The experimental values $\Delta z$ obtained in Ge and InP are compared in Fig.~\ref{fig:FigSMarburger} to calculations using the Marburger formula where power losses are [Eqs.~\eqref{eq:Marburger}, \eqref{eq:znl} and \eqref{eq:P(z)}] and are not [Eqs.~\eqref{eq:Marburger} and \eqref{eq:znl} for $P=P_{0}$] accounted for. Here, $\Delta z=z_{\rm{max}}-z_{g}$, where $z_{\rm{max}}$ corresponds to the position for which the maximum fluence $F_{\rm{max}}$ is reached, and $z_{g}$ corresponds to the geometrical focus. When power losses are not taken into account, the traditional Marburger formula can be used to estimate $\Delta z$ for about one order of magnitude above the critical power. However, this simple approach catastrophically fails to reproduce the experimental trends for higher power values. In contrast, our approach where power losses are accounted for allows us to fit the experimental data over several orders of magnitude. It is worth noting the applicability of our method for different multi-photon absorption orders (2PA and 3PA for Ge and InP, respectively). This approach allows us to determine $\beta_{2}^{\rm{eff}}$ and $\beta_{3}^{\rm{eff}}$ shown in the Primary Manuscript, Fig.~\ref{fig:Fig3}(c).\\

\begin{figure}[!ht]
\centering
\includegraphics[width=\linewidth]{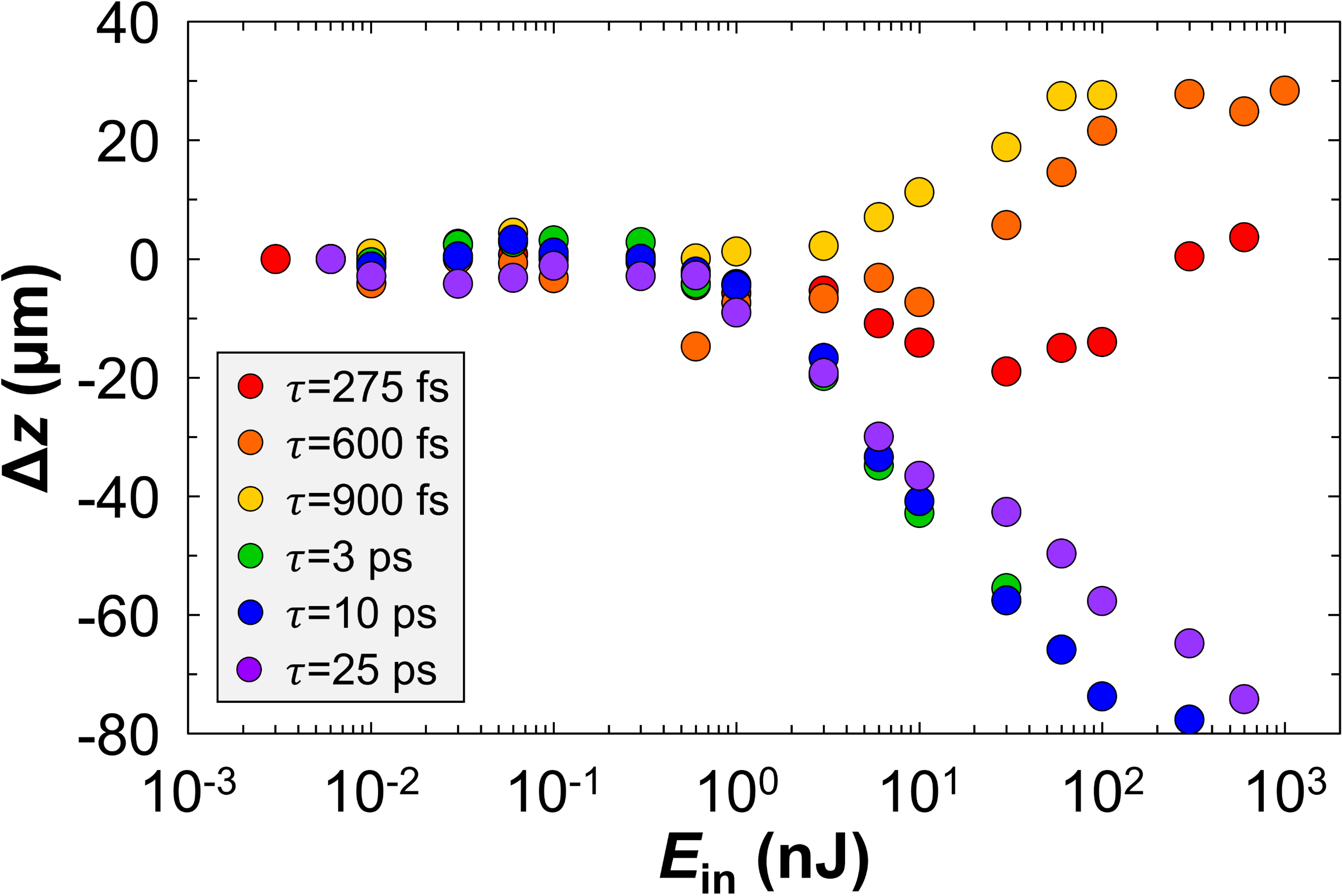}
\caption{\label{fig:FigSdeltazGe} Evolution of the nonlinear focal shift $\Delta z$ in Ge as a function of the input pulse energy $E_{\rm{in}}$ for various pulse durations $\tau$.}
\end{figure}

To apply the method we propose for determining the effective multi-photon absorption coefficient, a necessary condition is $\Delta z<0$, i.e., the focus shifts upstream the laser with respect to the geometrical focus. Moreover, $|\Delta z|$ must monotonically increase with $E_{\rm{in}}$. While these conditions are satisfied for the wide majority of our measurements, we noticed that these conditions are not fulfilled for Ge when employing sub-picosecond pulses. As shown in Fig.~\ref{fig:FigSdeltazGe}, the two conditions are fulfilled for pulse durations $\tau \ge 3$~ps. However, for $\tau=900$~fs, positive $\Delta z$ values are measured. This suggests that, in this regime, the assumption that multi-photon absorption does not affect propagation [Eqs.~\eqref{eq:Ed} and \eqref{eq:Ea}] is not valid anymore, and plasma defocusing may play an important role even for powers right above $P_{\text{cr}}^{\text{eff}}$. This is all the more confirmed when the pulse duration is further decreased to $\tau=600$ and 275~fs. In these high input intensity conditions, $\Delta z$ first decreases with $E_{\rm{in}}$, and then increases when $E_{\rm{in}}$ is further increased. This highlights the complexity of ultrafast laser-Ge interaction. As a direct consequence, $\beta_{2}^{\rm{eff}}$ cannot be extracted from data obtained in these conditions.\\

\end{document}